\documentclass{emulateapj}
\usepackage{rotating}

\usepackage{subfigure}
\usepackage{floatflt,graphicx}
\usepackage{lscape}

\newcommand{\HI}{H\,{\sc i}\ }
\newcommand{\HII}{H\,{\sc ii}\ }
\newcommand{\Ht}{H$_{2}$\ }
\newcommand{\arcs}{\arcsec }

\slugcomment{Accepted to ApJ: 27 November 2009}

\shorttitle{H$_2$ Line-cooling in Stephan's Quintet}
\shortauthors{Cluver et al.}

\begin{document}

%\title{H$_2$ Line-cooling in Stephan's Quintet : I- Mapping the Ghostly Outlines of Large-scale Turbulent Shocks}   

\title{Powerful H$_2$ Line-cooling in Stephan's Quintet : I - Mapping the Significant Cooling Pathways in Group-wide Shocks}

\author{M.E. Cluver\altaffilmark{1},\email{mcluver@ipac.caltech.edu} P.N. Appleton\altaffilmark{2,3}, F.~Boulanger\altaffilmark{4}, P. Guillard\altaffilmark{4},  P.~Ogle\altaffilmark{1}, P.-A.,~Duc\altaffilmark{5}, N.~Lu\altaffilmark{2}, J.~Rasmussen\altaffilmark{6}, W.T.~Reach\altaffilmark{1}, J.D.~Smith\altaffilmark{7}, R.~Tuffs\altaffilmark{8}, C.K.~Xu\altaffilmark{2}, M.S.~Yun\altaffilmark{9}}

\altaffiltext{1}{Spitzer Science Center, IPAC, California Institute of Technology, Pasadena, CA 91125}
\altaffiltext{2}{NASA Herschel Science Center, IPAC,  California Institute of Technology, Pasadena, CA 91125}
\altaffiltext{3}{Visiting Astronomer, Institute d'Astrophysique Spatiale, Universit\'{e} Paris Sud 11, Orsay, France}
\altaffiltext{4}{Institute d'Astrophysique Spatiale, Universite Paris Sud 11, Orsay, France}
\altaffiltext{5}{Laboratoire AIM, CEA/DSM - CNRS - Universit\'{e} Paris Diderot, DAPNIA/Service d'Astrophysique, CEA/Saclay, F-91191 Gif-sur-Yvette Cedex, France}
\altaffiltext{6}{Carnegie Observatories, 813 Santa Barbara St., Pasadena, CA 91101, Chandra Fellow}
\altaffiltext{7}{Department of Physics and Astronomy, Mail Drop 111, University of Toledo, 2801 West Bancroft Street, Toledo, OH 43606}
\altaffiltext{8}{Max-Planck-Institut f\"{u}r Kernphysik, Saupfercheckweg 1, 69117 Heidelberg,Germany}
\altaffiltext{9}{Department of Astronomy, University of Massachusetts, Amherst, MA}

\begin{abstract}

We present results from the mid-infrared spectral mapping of Stephan's
Quintet using the {\it Spitzer Space Telescope}\footnotemark[10]\footnotetext[10]{This work is based on observations made with the
{\it Spitzer} Space Telescope, which is operated by the Jet Propulsion Laboratory, California Institute of Technology under NASA contract 1407. Support for this work was provided by NASA through an award issued by JPL/Caltech.}. A 1000\,km\,$\rm{s^{-1}}$ collision ($t_{col}=5\times10^{6}$\,yr) has produced a group-wide shock and for the first time
the large-scale distribution of warm molecular hydrogen emission is revealed, as well
as its close association with known shock structures. In the main
shock region alone we find 5.0~$\times$10$^{8}$\,M$_{\odot}$ of warm \Ht
spread over $\sim$\,480\,kpc$^2$ and additionally report the discovery of
a second major shock-excited \Ht feature, likely a remnant of previous
tidal interactions. %, producing strong pure rotational \Ht emission.  
This brings the total \Ht line luminosity of the group in
excess of ${10}^{42}\, {\rm erg\, s^{-1}}$. In the main shock, the \Ht
line luminosity exceeds, by a factor of three, the X-ray luminosity from
the hot shocked gas, confirming that the H$_2$-cooling pathway dominates over the X-ray.
%Such abundant \Ht emission requires
%dust to survive in dense clumps in the postshock medium. 
$[$Si\,{\sc ii}$]$34.82\micron\
emission, detected at a luminosity of 1/10th of that of the H$_2$,
appears to trace the group-wide shock closely and in addition, we detect weak $[$Fe\,{\sc ii}$]$25.99\micron\ emission 
from the most X-ray luminous part of the shock. 
Comparison with shock models reveals that this emission is consistent with regions of fast shocks ($100 <V_{s} < 300$\,km\,$\rm{s^{-1}}$) experiencing depletion of iron and silicon onto dust grains. 

%Its strength relative
%to the \Ht suggests that it is collisionally excited in dense clumps
%associated with the molecular gas, rather than being a diffuse
%low-density shocked component. 

%The collisional excitation of
%$[$Si\,{\sc ii}$]$ in these clumps can account for the emission
%detected.  
Star formation in the shock (as traced via ionic lines,
PAH and dust emission) appears in the intruder galaxy, but
most strikingly at either end of the radio shock. The shock ridge
itself shows little star formation, consistent with a model in which
the tremendous \Ht power is driven by turbulent energy transfer from
motions in a post-shocked layer which suppresses star formation.  The
significance of the molecular hydrogen lines over other measured sources of
cooling in fast galaxy-scale shocks may have crucial implications for the
cooling of gas in the assembly of the first galaxies.

\end{abstract}

\keywords{galaxies: evolution -- galaxies: individual (NGC 7318b; NGC7319) -- galaxies: interactions -- galaxies: intergalactic medium -- physical data and processes: shock waves}

\section{Introduction}

Stephan's Quintet (hereafter SQ) 
%remains widely studied some 132 years
%after its discovery and is arguably the most famous of compact
%groups. The original group, as determined by \citet{Step77}, consisted
%of NGC 7317, 7318a, 7318b, 7319 and 7320. NGC 7320 was, however,
%subsequently shown to be a foreground galaxy with a largely discrepant
%redshift compared to the other four \citep{All70}. The other galaxies
is a strongly interacting compact group which has produced a highly disturbed
intragroup medium (IGM) \citep{Xu03,Xu05} through a 
%the current association is
complex sequence of interactions and
harrassment \citep{All80,Will02}.
This interplay has produced a
%between the members of the
%group was the discovery of a 
large-scale intergalactic shock-wave,
first observed as a narrow filament in the radio
continuum~\citep{All72}, and subsequently detected in the
X-ray~\citep{Piet97,Trin05,Osul09}.  The high-velocity
($\sim$1000\,km~s$^{-1}$) collision of an intruder galaxy, NGC 7318b,
with the intergalactic medium of the group~\citep{Sul01,Xu03} is
believed to be responsible for the shock-heating of the X-ray emitting
gas. Optical emission line ratios and observed broad linewidths provide
evidence that the region is powered by strong shocks and not star formation (Xu et al. 2003; Duc et al. 2010, in preparation).

The main elements of Stephan's Quintet are shown in Figure 1. Central to the system is the primary shock, as defined by the 20\,cm radio continuum emission. NGC 7318b, the intruder galaxy, lies to the West of the shock, while the large Seyfert 2 galaxy \citep{Dur94}, lies to the East. Other members of the group are also indicated. The peculiar extranuclear star formation region, named SQ-A \citep{Xu99,Xu03}, lies at the extreme northern end of the main shock wave. NGC 7318a (west of NGC 7318b) is also a strongly interacting group member, with NGC 7317 further away from the core.
% and is prominently detected at the infrared \citep{Xu99,Xu03} and ultraviolet wavelengths \citep{Xu05}.

%while the Seyfert 2 galaxy NGC 7319 \citep{Dur94}, lies to the East. %To the East is NGC 7319, a Seyfert 2 galaxy \citep{Dur94}.% with two lobes that are asymmetrically distributed along the minor axies of the galaxy \citep{Xan04}, but with no well-collimated jet. 
%The extranuclear star formation region, SQ-A, lies to the North of the primary shock and is prominently detected at the infrared \citep{Xu99,Xu03} and ultraviolet wavelengths \citep{Xu05}.
%Its UV-derived star formation rate (SFR) is 1.3\,$M_{\sun}\, \rm{yr^{-1}}$, consistent with a model of preshock Giant Molecular Clouds (GMCs) compressed by the surrounding shock gas that has formed from the \HI gas present before the collision \citep{Xu03}. 
%West of the shock lies NGC 7318b, the intruder galaxy believed to be colliding with the IGM.% and thus causing the formation and shock-heating of \Ht.

\begin{figure*}[!t]
\begin{center}
\includegraphics[width=15cm]{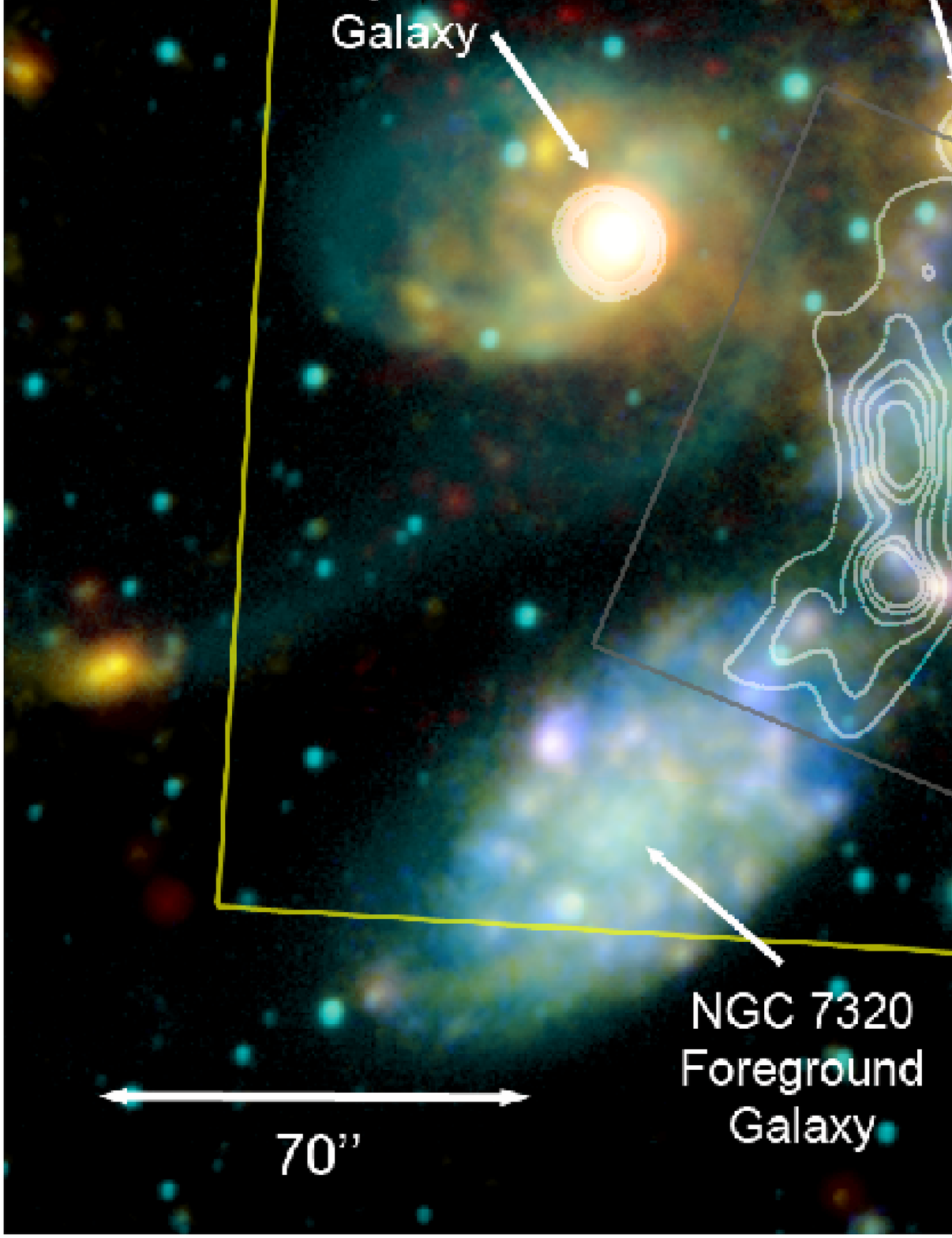}
\caption[]
{\small{Schematic composite image of Stephan's Quintet consisting of a {\it Galex} Near-UV image (blue), an optical R-band image (cyan) from \citet{Xu99}, an IRAC 8\micron\ image (yellow) and an image from the 16\micron\ IRS Peak-up Array (red). The contours represent the 20\,cm emission associated with the shock as obtained by the VLA \citep{Xu03}. The boundaries of the {\it Spitzer} IRS Spectral Mapping coverage (this paper) are shown as boxes. A color version of this image is available in the online publication.}}
\label{fig:int}
\end{center}
\end{figure*}

%**intro figure** shows combination of wavelengths and marks primary shock, SQA, bridge and Seyfert gal.
%radio contours -- X-ray front + obs outline (similar to the press release image)

The unexpected discovery of extremely
powerful, pure-rotational H$_2$ line emission from the center of
the shock \citep{App06}, using the {\it Spitzer Space Telescope} \citep{Wer04}, has sparked intense interest in the SQ system. 
Warm molecular hydrogen 
emission was found with a line luminosity exceeding the
X-ray luminosity from the shock. %by a large factor, thus becoming one of the
%important coolants in the shock. In addition, 
The mid-infrared (MIR) H$_2$ linewidth was resolved
($\sigma =$ 870 km s$^{-1}$) suggesting that the H$_2$-emitting clouds
carry a large bulk-kinetic energy, tapping a large percentage of the
energy available in the shock. 
A recent model of SQ, involving the collision between two inhomogeneous gas flows, describes H$_2$ formation out of the multiphase, shocked gas, and an efficient cooling channel for high-speed shocks as an alternative to X-ray emission \citep{Pg09}
%A recent model of SQ involving the
%interaction of a large-scale shock with a clumpy multi-phase medium
%describes a new (molecular) cooling channel for high-speed shocks as an alternative to X-ray emission \citep{Pg09}.

%The discovery of strong H$_2$ emission from SQ was followed soon after
%The publication of strong H$_2$ emission from SQ was quickly followed
%by new results for other galaxies which also show similar
%characteristics. 

Since the SQ detection, several other systems exhibiting similarly strong H$_2$ emission have been discovered.
\citet{Og07,Og09} find that a large-subset of the
local 3CR radio galaxies have extremely dominant MIR rotational
H$_2$ lines, often seen against a very weak thermal continuum. The low AGN and star formation power are insufficient to drive the MIR \Ht emission. Mechanical heating driven by the radio jet interaction with the host galaxy ISM is the favoured mechanism.
%indicate an alternative excitation mechanism generating the high H$_2$ line luminosities (10$^{41}$ $<$ L$_{\rm H_{2}}$ $<$ 5 $\times$ 10$^{42}$ ergs s$^{-1}$).
%do not appear to be capable of generating the high H$_2$ line
%luminosities
%These
%galaxies, termed Molecular Hydrogen Emission-line GalaxieS
%(MOHEGS, \citet{Og07}), appear to be powered primarily by shocks, since
%both the low-powered AGN continuua and very low star formation rates
%do not appear to be capable of generating the high H$_2$ line
%luminosities (10$^{41}$ $<$ L$_{\rm H_{2}}$ $<$ 5 $\times$ 10$^{42}$ ergs
%s$^{-1}$) observed. 
In addition, total H$_2$ luminosities in the range $10^{41} - 10^{43}$
ergs s$^{-1}$ have been detected in some central cluster
galaxies out to $z\sim 0.3$ \citep[][G. de Messieres - University of Virginia, Private communication]{Eg06} and in filaments in clusters \citep{Hat05}. The study of nearby
prototypes may provide valuable insight into the nature of these more
distant systems.  The large scale ($\sim$~30~kpc) of the SQ
shock is well-suited for such a study.

In this Paper I, we extend the single pointing observations of
\citet{App06} to full spectral maps of SQ using the IRS instrument on the {\it Spitzer} Space Telescope, hereafter {\it Spitzer}.
%This shows, for the first time,
%the astonishing extent and spatial distribution of warm H$_2$ and
%places the emission in the context of the known shocked X-ray
%gas and the compact group as a whole.  The new observations have
%revealed the detection and distribution of $[$Fe\,{\sc ii}$]$25.98\micron\ 
%and $[$Si\,{\sc ii}$]$34.81$\mu$m
%emission from the shock, as well as PAH (polycyclic aromatic hydrocarbon) and dust emission from
%star forming regions. 
In Paper II, we will present detailed
2-dimensional excitation maps of the H$_{2}$ emission across the face
of the X-ray emitting shock, and compare them with models.
In addition, several other papers are being prepared by
our team which will discuss the relationship between the UV/X-ray
emission and emission from dust.

In Section 2, we present our observations and data reduction methods.
In Section 3 we discuss the mapping results for various detected lines. In Sections 4, 5 and 6 we present our discussion on the properties of the system and in Section 7 the implication for galaxy formation. In Section 8 we present our
conclusions. Additional material is included as appendices, with a discussion on NGC 7319 in Appendix A, and a reanalysis of the high resolution MIR spectrum of the shock, as well as renanalysis of the X-ray data presented in Appendix B and C, respectively.

Throughout this paper, we assume a systemic velocity of $v = 6600\, {\rm
  km\, s^{-1}}$ for the group, corresponding to a distance of 94 Mpc
with $H_{\rm o}=70\, \rm{km}\, \rm{s}^{-1} Mpc^{-1}$.

\section{Observations and Data Reduction}

Mid-infrared spectroscopy of the shock region in SQ was obtained using
the IRS instrument \citep{Hou04} onboard {\it Spitzer}.
Observations were done in low-resolution mapping mode,
using the Short-Low ($R\sim60-127$; $5.2-14.5$ \micron) and Long-Low
($R\sim 57-126$; $14.0-38.0$ \micron) modules and taken on January 11
2008 and December 10 2007, respectively. Figure 1 displays the outline of
the areas observed superimposed on a composite image of the group. 

The SL spectral mapping consists of two separate, partially
overlapping maps, centered North and South on the X-ray emission
associated with the shock. The map was constructed with 23 steps of 2.8\arcs
(0.75 $\times$ slit width) perpendicular to the slit and one parallel step of
7.2\arcs. Observations consisted of 60\,s integrations with 5 cycles
per step.

The LL module was used to map an area of $\sim 2.8\arcmin \times
3.2\arcmin$ using 21 steps of 8.0\arcs\ (0.75 $\times$ slit
width) perpendicular to the slit and a parallel step of 24.0\arcs. An
integration time of 120\,s was used with 3 cycles per step.

Primary data reductions were done by the {\it Spitzer} Science Center
(SSC) pipeline, version S17.0.4 and S17.2.0 for SL and LL respectively,
which performs standard reductions such as ramp fitting, dark current
subtraction and flat-fielding. Background subtraction for LL data was performed by
subtracting dedicated off-source observations, with the
same observing mode, taken shortly after the mapping sequence. In
the case of SL, where for scheduling reasons the dedicated ``off''
observations were too far away in time to be optimal, backgrounds were generated from
observations at the periphery of the map that contain no spectral
line signatures. 

Examination of the pipeline products showed that the stray-light-corrected images (that account for potential spillover from the
peak-up arrays onto the SL1/2 spectral apertures) contained wavelength-dependent
over-corrections in some of the images, seemingly due to a high background of
cosmic rays during the SL portion of the mapping. To negate this effect, the alternative flat-fielded images (also available from the science pipeline) that are uncorrected for crosstalk and straylight removal were used.
%Unlike standard pipeline processing, we did not perform
%the stray-light correction to account for potential spillover from the
%peak-up arrays onto the SL1/2 spectral apertures. This was due to the
%stray-light-corrected images containing wavelength-dependent
%over-corrections in some of the images due to a high background of
%cosmic rays during the SL portion of the mapping. 
It was determined that after background subtraction, any uncorrected stray
light in the SL spectral extraction areas was unmeasurable at the $<
2-3$\% level. Thus stray-light correction was unneccesary, and the
resulting Basic Calibrated Data (BCD) images used were of better quality
than the standard BCDs.
 
For all modules, individual frames for each pointing were
median-combined and obvious ``bad'' pixels were replaced using customised
software that allows for manual ``average'' replacement.  The spectra were
assembled into spectral cubes for each module using the software tool,
CUBISM \citep{Smi07}. Further bad pixel removal was performed within
CUBISM. Spectral maps were generated by making continuum maps on
either side of a feature and subtracting the averaged continuum map
from the relevant emission line map. One dimensional spectra were further 
extracted from the data cubes using matched apertures.

Broad-band images at 16 and 24$\mu$m were obtained with the {\it
  Spitzer} IRS blue Peak-up Imager (PUI) and the MIPS instrument. The
PUI was obtained in a 5 x 5 map with 3 cycles of 30\,s duration each on
2007 December 10.  The MIPS instrument \citep{Riek04} on {\it Spitzer}
obtained 24\micron\ imaging of SQ on 2008 July 29, achieving a spatial
resolution of $\sim$6\arcs. Primary data reduction was done by the
{\it Spitzer} Science Center (SSC) science pipeline (version S18.0.2)
run through the MOPEX software, and for the MIPS image a smooth 2D
polynomial background was further removed to correct for a large-scale
background gradient.
{\it Spitzer} IRAC 3.6, 4.5, 5.8 and 8.0 \micron\ data of SQ (P.I. J.R. Houck) were obtained from the SSC archive; these were reduced using science pipeline version S18.0.2. The final mosaics have a pixel scale of 0.61\arcs.

\section{Results}

\subsection{The Molecular Hydrogen Distribution}\label{molhy}

The pure rotational transitions of molecular hydrogen can be excited
by several mechanisms. These include FUV (Far Ultraviolet) induced pumping, and possible additional collisional heating, of the \Ht in photodissociation
regions associated with star formation (e. g. Black \& van Dishoeck 1987,
Hollenbach \& Tielens 1997), hard X-rays penetrating and heating
regions within molecular clouds, which in turn excite \Ht via
collisions with electrons or hydrogen atoms (Lepp \& McCray 1983;
Draine \& Woods 1992) and finally collisional excitation of \Ht due to
acceleration produced by shocks (e.g. Shull \& Hollenbach 1978). The pure rotational
MIR line ratios are not especially good diagnostics for distinguishing
between these mechanisms since all three mechanisms discussed  can lead to 
well thermalized level distributions of lower-level rotational states. 
The rotational \Ht emission lines do, however, allow us to 
trace gas at different temperatures and compare with model predictions (this will be the main
emphasis of Paper II). Higher level transitions 0-0 S(3)-S(5) tend to trace warmer gas,
whereas the S(0) and S(1) lines are sensitive to the ``coolest'' warm H$_{2}$.

\begin{figure*}[!p]
\begin{center}
\includegraphics[width=13.5cm]{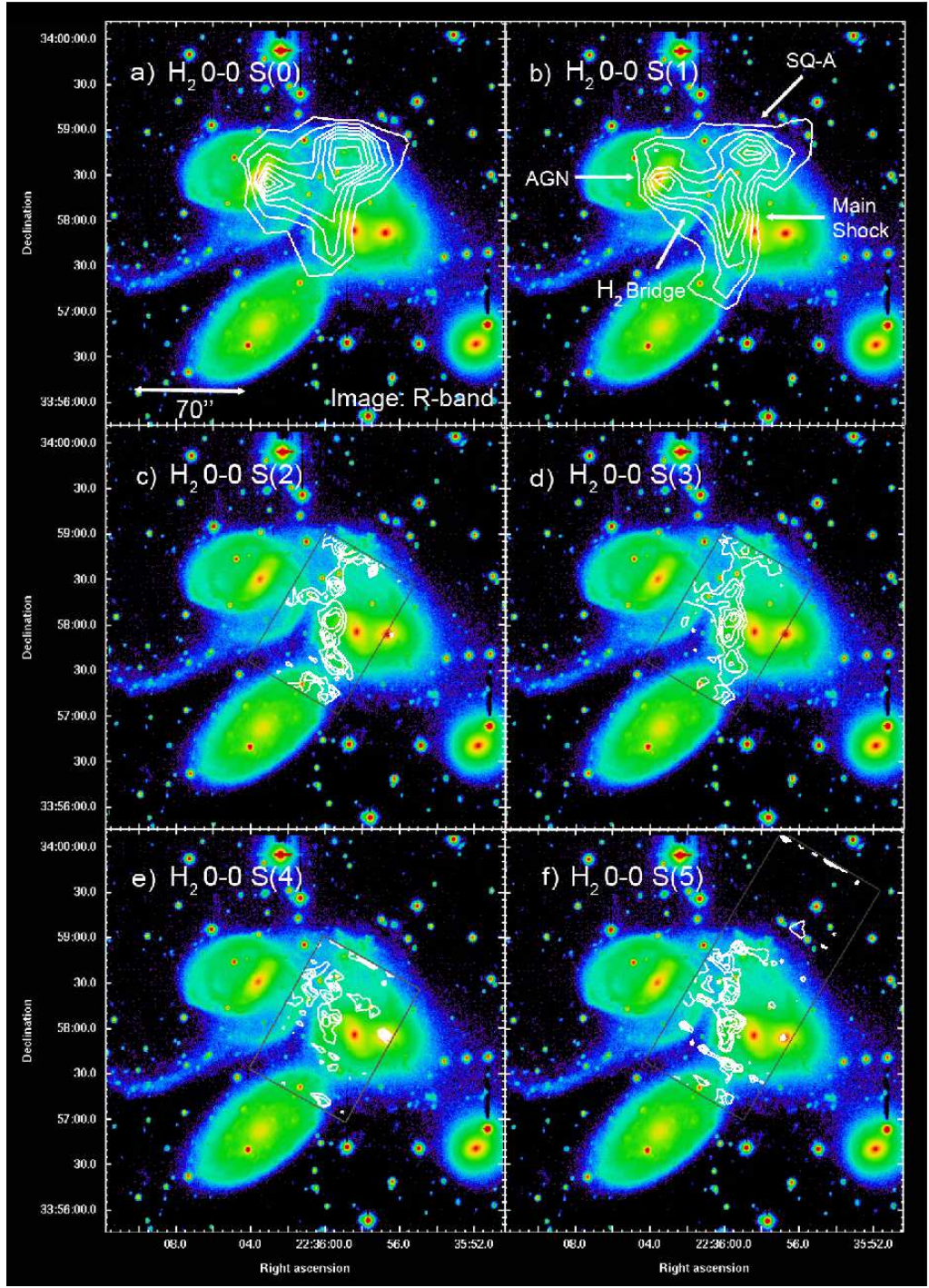}
\caption[The large-scale distribution of warm \Ht.]
{\scriptsize{The large-scale distribution of warm \Ht in Stephan's Quintet with each detected 0-0 transition overlaid on an $R$-band image from \citet{Xu99}. The grey box indicates the limit of the SL module coverage. Contour levels are as follows (all in units of MJy/sr): (a) 0.1, 0.14. 0.19, 0.28, 0.32, 0.37, 0.41, 0.46, 0.5, (b) 0.3, 0.53, 0.75, 0.98, 1.2, 1.43, 1.65, 1.88, 2.1, (c) 0.3, 0.4, 0.5, 0.6, 0.7, (d) 0.25, 0.42, 0.58, 0.75, 0.92, 1.08, 1.25, (e) 0.11, 0.19, 0.27, 0.36, 0.44, 0.52, 0.6 and (f) 0.3, 0.46, 0.61, 0.77. 0.93, 1.09, 1.24, 1.4. The SL coverage for (f) is larger due to incorporating the ``parallel mode'' SL2 slit observations from the ``off-target'' positions when SL1 mapping observations were the primary observing mode.}}
\label{fig:opt_h}
\end{center}
\end{figure*}

Although the line ratios themselves cannot be used directly as shock diagnostics,
in Stephan's Quintet the distribution of large-scale X-ray and radio
emission, plus optical emission line diagnostics, provide strong
evidence that the giant filament seen in Figure 1 is the result of a
strong shock. In Appleton et al. (2006) this fact was used to reveal the surprising
association of detected \Ht emission with the shock. However, in this paper we
can make a more definitive association of the emission with the shock by means of spectral maps.
%stronger since we now have maps of the
%emission.

The spectral cubes were used to extract maps of all the pure rotational
emission lines of molecular hydrogen that were detected, namely the
0-0 S(0)\,28.22\micron, S(1)\,17.03\micron, S(2)\,12.28\micron, S(3)\,9.66\micron, S(4)\,8.03\micron\ and S(5)\,6.91\micron\ lines. Specific intensity contour maps of these
lines are presented in Figure 2 overlaid on a $R$-band image of the region from \citet{Xu99}. The S(0) and S(1) lines were mapped by the LL modules, while S(2) - S(5) transitions were 
%the remaining transitions In Fig. 2c - f we show the distribution of the \Ht transitions
mapped by the SL modules of the IRS. As indicated in Figure 1, the SL
observations were concentrated on the main shock to provide high
signal to noise (S/N) measurements there. As a result, these maps do not fully cover SQ-A or NGC 7319.
%Local backgrounds were subtracted on either side of each
%line to create the maps and remove any continuum. 
We note that the S(4) line at
8.03\micron\ (Fig. 2e) is faint, and also suffers from contamination %($\sim$~40\%) 
from the PAH
bands at 7.7\micron\ and 8.6\micron.
%{\bf how much contamination? how  did you reduce it in the map-phil?}

The contours indicate powerful, widespread emission running North-South
along the shock ridge (see Fig. 1). In addition we detect
strong emission from the star forming region, SQ-A, as well as
associated with NGC 7319. We discuss further details of the AGN-like
MIR emission lines from NGC~7319 in Appendix A.  

Figures 2a-d reveal a new \Ht structure running eastward from the ``main'' shock
ridge.  In what follows, we refer to this feature as the
``bridge''. This structure is observed faintly in the {\it Chandra}
\citep{Trin03} and XMM \citep{Trin05} X-ray images and detected as
faint H$\alpha$ emission by \citet{Xu03}, but is not strong in radio continuum
images. 
%The likely origin of this feature will be discussed later in
%the paper.

%The specific intensity contours for the H$_{2}$ S(0) line are shown in Fig. 2a and
%indicate 
%powerful, widespread emission running North-South along the
%direction of the shock shown in Fig. 1. In addition we also see
%strong emission from the star forming region SQ-A, as well as
%associated with the AGN in NGC 7319 \citep{Xan04}. A further
%surprise in this system is the discovery of a new \Ht structure running
%from the ``main'' shock in an eastward direction towards NGC 7319. In
%what follows, we refer to this feature as the ``bridge''. This feature
%is seen faintly in the Chandra \citep{Trin03} and XMM \citep{Trin05} X-ray images and 
%detected as a faint H$\alpha$ ``bridge'' by \citet{Xu03}. 

As is evident in Figure 2, there is distinct variation in the distribution of the warm \Ht emission.
The brightest 0-0~S(0) emission (Fig. 2a)  appears to be concentrated towards the
North of the shock, whereas the 0-0~S(1) transition emission (Fig. 2b) appears %with emission extended across and along the shock front,
more concentrated towards the center.  
%Compared to the S(0) emission,
%the S(1) emission appears to be more evenly distributed along the main
%shock structure. 
%The ``bridge'' structure noted above is
%strongly detected. 
The S(0) and S(1) maps demonstrate that
the \Ht in the bridge terminates in a large clump a few arcsecs west
of the nucleus of the Seyfert 2 galaxy NGC~7319, and in a small detour
to the North (especially in the S(1) map which has the highest
S/N). 
%These two features near NGC 7319 are exactly correlated with two
%major concentrations of CO emission detected by \citet{Gao00}.

The S(2) through S(5) lines clearly show that the warm \Ht emission
breaks into clumps in the shock. Despite the limited coverage compared
to the LL mapping, the base of the ``bridge'' is visible and SQ-A is
partially covered.  SQ-A is fully covered by the SL2 module (because
of fortuitous ``off observation'' coverage) and hence the S(5)
emission line reveals that in SQ-A the \Ht emission is also clumpy.

The %high signal S/N 
molecular-line maps provide considerable information about the excitation of
%There is considerable information in these maps which provide high
%signal to noise information about the excitation of 
the \Ht along and
across the X-ray shock, but these discussions will be deferred to a
full 2-D modelling of the \Ht excitation in Paper II (Appleton et
al., in preparation). Instead we shall limit ourselves to global
properties of the \Ht here. In Section \ref{spec} we shall present spectra of some
selected regions of the emission and discuss a global excitation
diagram for the shock.

%\begin{figure*}[!p]
%\begin{center}
%\includegraphics[width=15cm]{opt_comb.eps}
%\caption[]
%{\small{b) Contour levels are at 0.1, 0.14. 0.19, 0.28, 0.32, 0.37,
%0.41, 0.46, and 0.5 MJy/sr.  c) Contour levels are at 0.3, 0.53,
%0.75, 0.98, 1.2, 1.43, 1.65, 1.88 and 2.1 MJy/sr. d) Contour levels
%are at 0.3, 0.4, 0.5, 0.6, 0.7 MJy/sr. e) Contour levels are at 0.25,
%0.42, 0.58, 0.75, 0.92, 1.08 and 1.25 MJy/sr. f) Contour levels are
%at 0.3, 0.46, 0.61, 0.77. 0.93, 1.09, 1.24, 1.4 MJy/sr.}}
%\label{fig:opt_h}
%\end{center}
%\end{figure*}

\begin{figure*}[t!]
\begin{center}
\includegraphics[width=16cm]{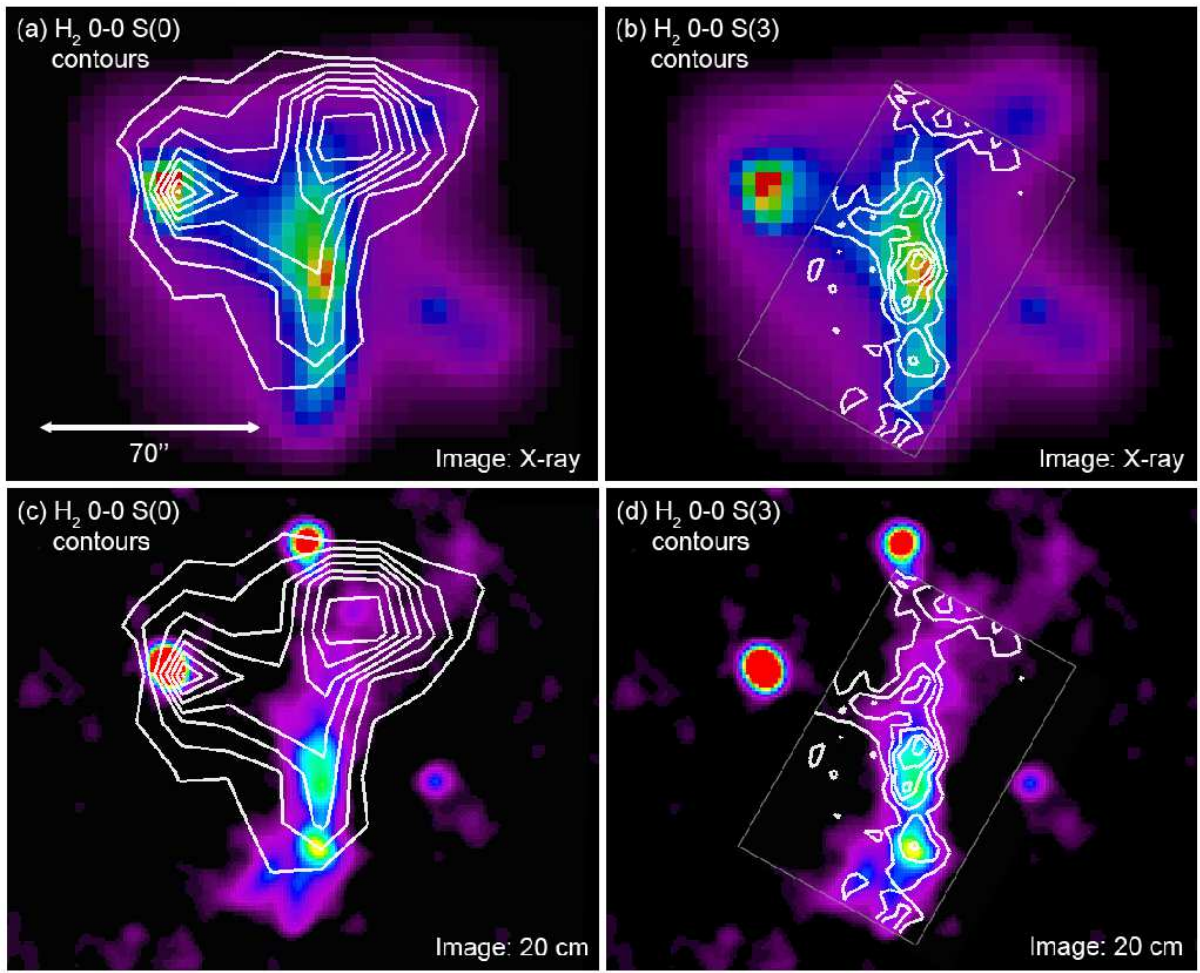}
\caption[] {\small{ X-ray image from XMM-Newton \citep{Trin05} in the 1.5$-$2.5 keV
band with (a) \Ht S(0) contours and (b) \Ht S(3) contours
overlaid. Also shown are 1.4 GHz radio continuum images from the VLA \citep{Xu03} with (c) \Ht S(0) contours and (d) \Ht S(3)
contours overlaid. Contours levels are as in Fig \ref{fig:opt_h}. }}
\label{fig:xray_rad}
\end{center}
\end{figure*}

To further demonstrate the close connection between the \Ht emission
and the main global shock-wave in SQ, we now consider the distribution
of warm \Ht in relation to the X-ray and radio emission.  Figure
\ref{fig:xray_rad}a and b show the S(0) and S(3) contours overlaid on an
XMM-Newton X-ray image \citep{Trin05} of
Stephan's Quintet. The warm molecular
hydrogen is distributed along the length of the main North-South (NS)
X-ray shock ridge and along the ``bridge'', demonstrating the remarkable
projected coexistence of of hot X-ray plasma (10$^{6}$ $<$ T $<$ 10$^{7}$ K) and warm H$_{2}$ (10$^{2}$ $<$ T$<$ 10$^{3}$ K). Although the \Ht appears to follow the X-ray,
there are subtle differences. The cooler S(0) line has
emission concentrated to the North and follows the X-ray less closely
compared to the warmer S(3) line. The S(3) line exhibits a clear
correspondence to the X-ray, notably at the center of the shock, where
we find peaks at both wavelengths. Thus the region of greatest shock heating, as traced by the stronger X-ray emission, appears to correspond to the higher-J \Ht transitions, perhaps implying a causal connection. The
intergalactic star formation region SQ-A, is essentially absent in
X-ray emission, as seen in Figure \ref{fig:xray_rad}a, but is strongly
detected in H$_2$.

We find a similar picture in the radio continuum (Fig. \ref{fig:xray_rad}c \& d)
with the S(0) line demonstrating correspondence with the main shock, but
dominated by emission in the North where we observe less powerful radio
emission. The S(3) (and S(2)--see Fig. 2c) line presents a much tighter
correlation with regions of the shock that are more radio luminous
than the lower-J transitions. 
%However, the correlation breaks down for
%the warmest gas which would be represented by the S(4) and S(5)
%lines. 
The radio emission is quite likely sensitive to the most
compressed regions of the shock where cosmic ray particles are
accelerated more strongly \citep{Hel03,App99}, whereas the brightest
X-ray patches are likely due to the fastest regions of the shock
\citep{Pg09}. As already mentioned the ``bridge'' emission is detected only
faintly in the X-ray and is weak or absent at radio continuum wavelengths. This noticeable difference compared to the main shock likely implies that the conditions that give rise to strong synchrotron emission in the main shock are absent in the bridge.

%could suggest
%older shock heating in the ``bridge'' region.

%We will discuss the potential significance of the \Ht distributions in Section \ref{glob_prop}.

\subsection{Shock-related Fine-structure Line Emission}

\begin{figure*}[!th]
\begin{center}
\includegraphics[width=16cm]{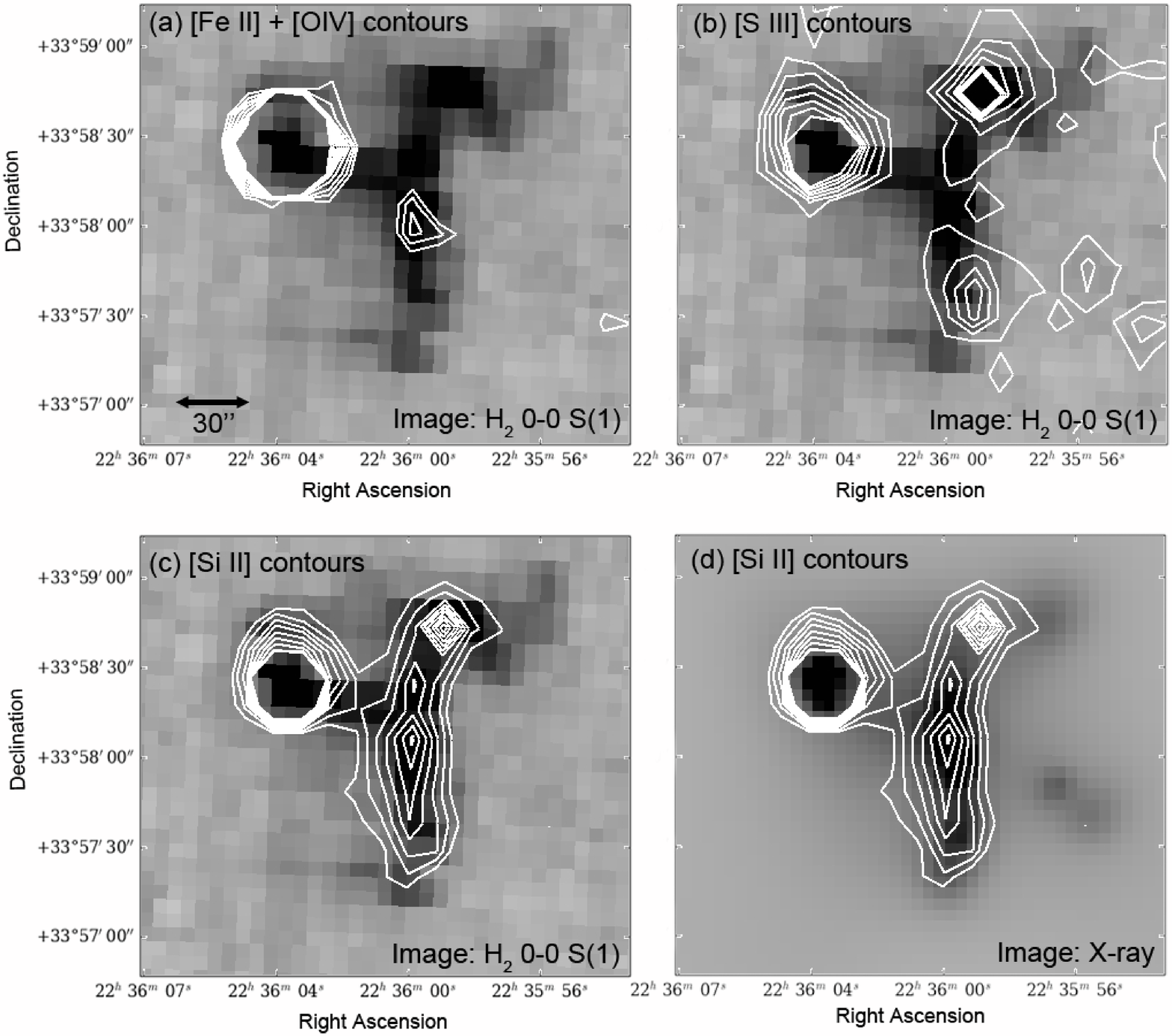}
\caption[]
{\small{Fine structure emission ($[$Fe\,{\sc ii}$]$25.99\micron\ and $[$O\,{\sc iv}$]$25.89\micron\ blend, $[$S\,{\sc iii}$]$33.48\micron\ and $[$Si\,{\sc ii}$]$34.82\micron) compared to the distribution of \Ht S(1) emission. Also shown are $[$Si\,{\sc ii}$]$34.82\micron\ contours compared to X-ray 0.5-1.5 keV emission. Contour levels are as follows (all in units of MJy/sr): (a) 0.08, 0.1, 0.13, 0.15, 0.165, 0.18, 0.2, 0.22, 0.23, 0.25, (b) 0.2, 0.33, 0.47, 0.6, 0.73, 0.87, 1.0 with (c) and (d) having levels of 0.5, 0.66, 0.83, 0.99, 1.15, 1.31, 1.48, 1.6, 1.8. These levels are chosen to best show intensity of emission associated with the shock and do not reflect the full intensity range for NGC 7319.}}
\label{fig:fs}
\end{center}
\end{figure*}

Emission from fine-structure lines provide key diagnostics that trace
the interplay between the various constituents of the shocked
interstellar medium. In \citet{App06} the spectra were
limited to the very core of the shock and only weak emission was
detected from all but two metal lines, namely $[$Ne\,{\sc ii}$]$12.81\micron\
and $[$Si\,{\sc ii}$]$34.82\micron\ (this data has been reanalysed and is presented in Appendix B).
%Here, however, we cover a much larger area and
%go deeper than the original IRS high resolution observations.
In this section we
discuss the spatial distribution of emission from the $[$Fe\,{\sc ii}$]$25.99\micron, $[$O\,{\sc iv}$]$25.89\micron, $[$S\,{\sc iii}$]$33.48\micron,
$[$Si\,{\sc ii}$]$34.82\micron, $[$Ne\,{\sc ii}$]$12.81\micron\ and $[$Ne\,{\sc iii}$]$15.56 \micron\ lines.

In Figure \ref{fig:fs} we present the specific intensity contours of the
$[$Fe\,{\sc ii}$]$25.99\micron\ and $[$O\,{\sc iv}$]$25.89\micron\ blend, 
$[$S\,{\sc iii}$]$33.48\micron\ and
$[$Si\,{\sc ii}$]$34.82\micron\ emission lines. 

Given the low spectral
resolution of the SL and LL modules of {\it Spitzer}, we cannot
distinguish between emission from $[$Fe\, {\sc ii}$]$25.99\micron\ and
$[$O\,{\sc iv}$]$25.89\micron. However, except in the direction of the
Seyfert II galaxy NGC~7319, the emission near 26\micron\ seen in
Figure \ref{fig:fs}a is likely to be pure $[$Fe\,{\sc ii}$]$ with
little contamination from $[$O\,{\sc iv}$]$ as there is no evidence
from the spectra of high-excitation emission from the intragroup
medium in SQ. For example, $[$O\,{\sc iv}$]$ has an excitation
potential of 56 eV (compared to 7.9 eV of $[$Fe\,{\sc ii}$]$), and yet
\citet{App06} have shown that the $[$Ne\,{\sc iii}$]$/$[$Ne\,{\sc ii}$]$ ratio 
is low suggesting low-excitation conditions for the ions in the shock, further supported in Section \ref{em}.
%{\bf[Reference to the Allen paper with Dopita?]}
Assuming that, apart from towards NGC~7319, $[$Fe\,{\sc ii}$]$ dominates the $[$Fe\,{\sc ii}$]$+$[$O\,{\sc iv}$]$ complex, we detect faint emission from $[$Fe\,{\sc ii}$]$25.99\micron\ %(more likely given its lower ionisation potential) 
at the location of the center of the shock (as defined by the X-ray ``hotspot'' in
Fig. \ref{fig:xray_rad}a). The energetic requirements for shocks to produce strong $[$Fe\,{\sc ii}$]$ 
emission are usually present in $J$ shocks while the ion
abundance in $C$ shocks are low in comparison \citep{Hol89}. We will discuss the production of $[$Fe\,{\sc ii}$]$ in the shock in Section 6.2.

The $[$S\,{\sc iii}$]$33.48\micron\ distribution is presented in Figure
\ref{fig:fs}b. This fine structure line acts as a strong tracer of
\HII regions \citep{Dal06} and we observe emission from SQ-A and from
other regions of star formation in the South (see Section
\ref{SF}). In the primary shock region the distributions
of $[$Fe\,{\sc ii}$]$ and $[$S\,{\sc iii}$]$ are anti-correlated.

In strong contrast to the weak $[$Fe\,{\sc ii}$]$ emission, there is copious
$[$Si\,{\sc ii}$]$34.82\micron\ emission (Fig. \ref{fig:fs}c), which follows the 
S(1) distribution closely with respect to the primary shock, as mapped by X-ray emission (Fig. \ref{fig:fs}d).

Although $[$Si\,{\sc ii}$]$ is commonly found in normal \HII regions,
we will demonstrate below that, apart from in SQ-A, the strong silicon emission
does not correlate with regions of strong PAH emission (tracing star formation) in SQ, but instead closely follows the \Ht and X-ray (shock) distributions. $[$Si\,{\sc ii}$]$ acts as an efficient coolant of X-ray-irradiated
gas and is predicted to be one of the top four cooling lines under
these circumstances \citep{Mal96}. We will discuss the excitation of $[$Si\,{\sc ii}$]$ in the shock in Section 6.2.

\begin{figure*}[!thp]
\begin{center}
\includegraphics[width=16cm]{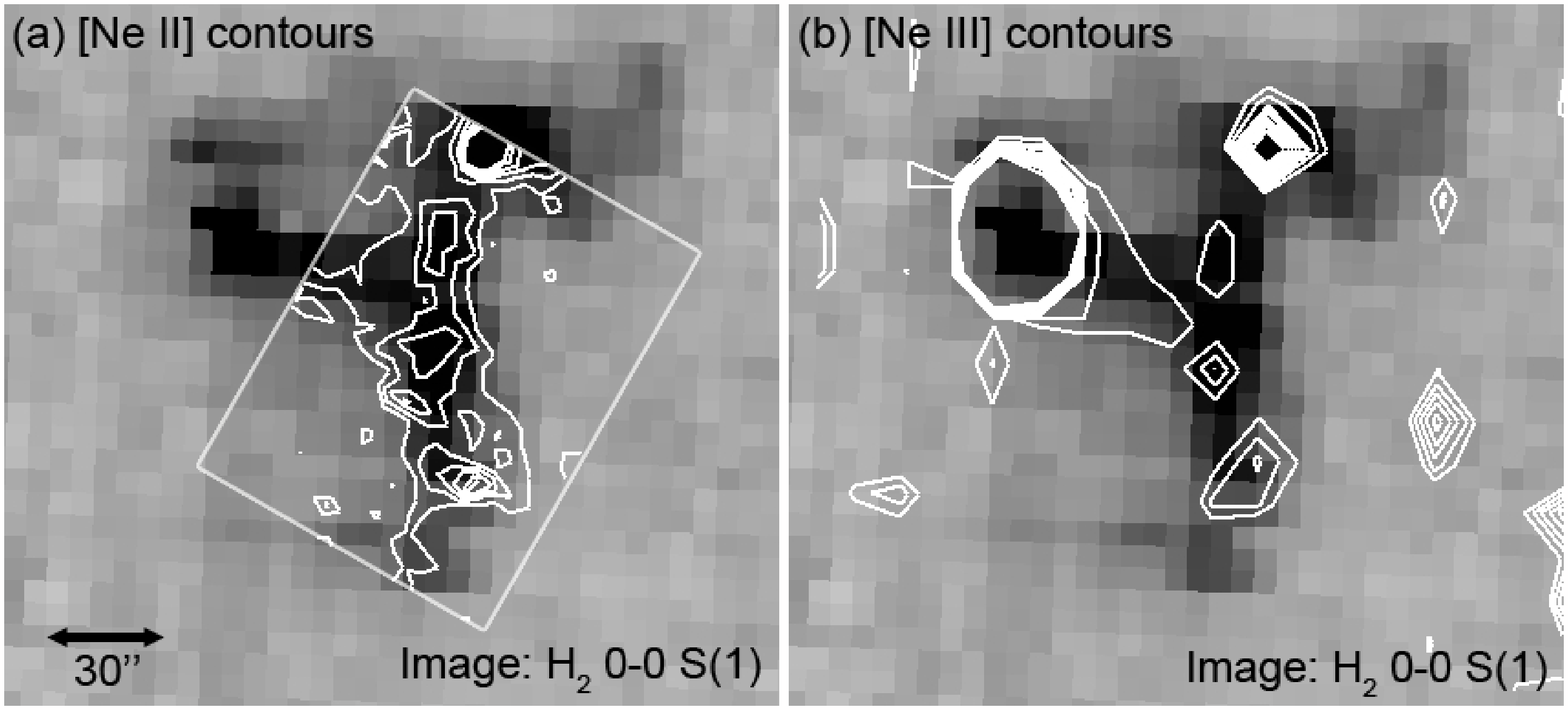}
\caption[]
{\small{$[$Ne\,{\sc ii}$]$12.81\micron\ and $[$Ne\,{\sc iiI}$]$15.56\micron\ emission compared to the distribution of \Ht S(1) emission. Contour levels are as follows (in units of MJy/sr): (a) 0.22, 0.4, 0.57, 0.75, 0.92, 1.1 and (b) 0.06, 0.08, 0.1, 0.13, 0.16, 0.18, 0.2, 0.23, 0.25.}}
\label{fig:ne}
\end{center}
\end{figure*}

$[$Ne\,{\sc ii}$]$12.81\micron\ (with an ionisation potential of 21.6 eV) is also represented in the shock, as shown in Figure \ref{fig:ne}a, although it is
also emitted from some \HII regions associated with star formation, such as SQ-A and the star-forming region South of NGC 7318b (see Section \ref{SF}). The $[$Ne\,{\sc iii}$]$15.56\micron\ contours (with an ionisation potential of 41 eV) are shown in Figure \ref{fig:ne}b and are associated with excitation in the shock, as well as from regions of star formation (see Section \ref{SF}).
%$[$Ne\,{\sc ii}$]$ and $[$Ne\,{\sc iii}$]$ are typical lines produced in \HII gas
%regions and have ionisation potentials of 21.6 eV and 41 eV
%respectively. 
The higher ionisation line of $[$Ne\,{\sc iii}$]$ 
is much weaker in the shock ridge, but regions of emission
correspond closely to peaks seen for the \Ht S(3) line. The
$[$Ne\,{\sc ii}$]$ emission, however, exhibits clearly extended emission
with regions of greatest luminosity matching those seen for the \Ht
S(3) line. The $[$Ne\,{\sc ii}$]$ emission suggests excitation from the shock, with $[$Ne\,{\sc iii}$]$ found at the location of the center of the shock, similar to what is observed for $[$Fe\,{\sc ii}$]$.

\subsection{Star Formation Tracers and Dust Emission in Stephan's Quintet}\label{SF}

\begin{figure*}[!thp]
\begin{center}
%\begin{minipage}{7.5cm}
%\begin{flushleft}
\includegraphics[width=16cm]{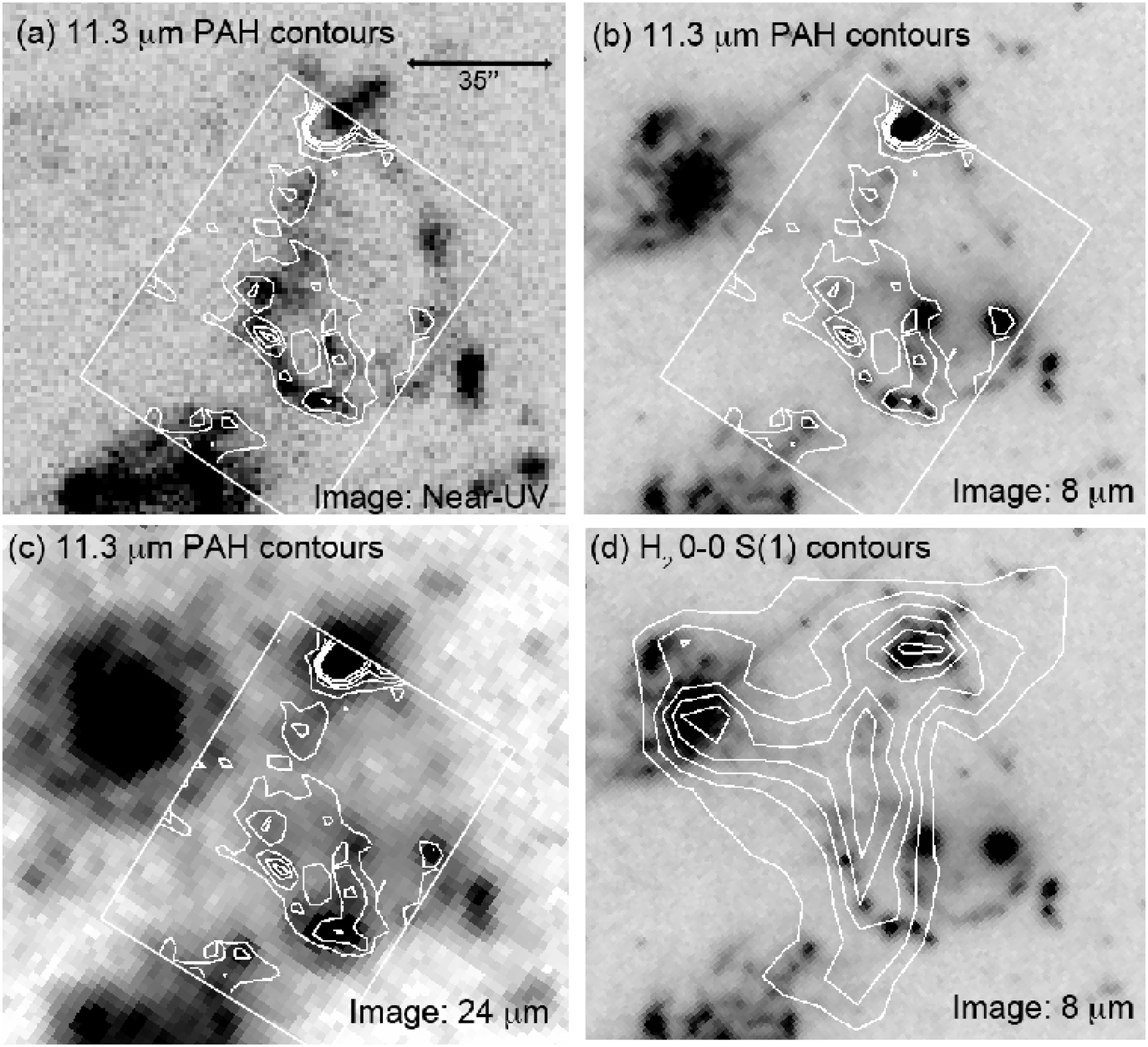}
\caption[]
{\small{Dust as tracers of star formation. The 11.3\micron\ PAH contours overlaid on (a) the Near-UV, (b) an 8\micron\ image and (c) a 24\micron\ image. By comparison the \Ht 0-0 S(1) contours show little correlation with the 8\micron\ image (as shown in d). Contour levels are as follows (in units of MJy/sr): (a), (b) and (c) are 0.2, 0.35, 0.5, 0.65, 0.8. Contour levels for (d) are as in  Fig. \ref{fig:opt_h}b.}}
%\label{fig:dust}
%\end{flushleft}
%\end{minipage}
\label{fig:dust}
\end{center}
\end{figure*}

%{\bf ***NeII, NeIII, SIII, SiII/SIII*** discuss here (no measurements yet)?}

Previous observations and spectroscopy of SQ \citep{Xu05} have determined that there
are some regions of star formation associated with the spiral arms
of the intruder galaxy NGC~7318b. We discuss in this section how these regions, which have
a different spatial distribution from the shocked gas, are correlated with the PAH emission
we detect in the IRS spectra. 

In Figure \ref{fig:dust}a there is a strong correlation between the
11.3\micron\ PAH distribution (from the IRS cube) superimposed on a
Near-UV image from {\it Galex} \citep{Xu05}, which maps the UV emission
from hot stars associated with weak star formation from the system. This correlation suggests that the PAH molecules are 
excited by star formation. A similar
close correlation is shown in Figure \ref{fig:dust}b where we
overlay the 11.3\micron\ contours on the IRAC 8\micron\ band, which is dominated by
the 7.7 and 8.6 $\mu$m feature. It is noticeable, however, that regions with strong 11.3\micron\ emission in the shock, do not appear similarly strong at 8\micron. This point will be addressed in Section 6.4.

We compare the 11.3\micron\ PAH map to the distribution of warm dust in Figure \ref{fig:dust}c, using the MIPS 24$\mu$m map of SQ. Again, it is clear that there is a good correlation between the PAH emission and the thermal dust, most of which seems only poorly correlated with the shock ridge. The lack of conspicuous star formation in the ridge was also observed by \citet{Xu03}. 

The main point we emphasize here is that the dust, PAH and UV
emission appears to be associated with previously known star formation
regions and no additional star formation is observed in the shock; this can be seen in Figure \ref{fig:dust}d
showing the H$_{2}$ 0-0 S(1) emission overlaid on an IRAC
8\micron\ image. There is little correspondence between the \Ht emission
in the shock and 8$\mu$m (hot dust plus PAH) image. This is important
because it implies that there is very little triggered star formation
in the molecular gas associated with the shock-excited H$_{2}$.

Figure \ref{fig:dust}c demonstrates that there is only faint dust emission at
24$\mu$m from the shock ridge. The presence of dust in the shock is
required in the model of Guillard et al. (2009) to explain the
formation of \Ht behind the shock and we do observe evidence of depletion onto dust grains (see Section 6.2). Thus the faint 24\micron\ emission could be the result of destruction of Very Small Grains (VSGs), with only larger grains surviving, or indicate that the grains are cold and radiating more strongly at
longer wavelengths, where {\it Spitzer} has the least spatial
resolution. A more detailed description of the
dust and faint PAH emission in the SQ group (including results from MIPS 70\micron\ imaging) is discussed in separate
papers (Guillard et al. 2010.; Natale et al. 2009, in preparation). Guillard et al. (2010) show that the IR emission in the shock is faint due to dust being heated by a relatively low intensity UV radiation field and determine a Galactic PAH/VSG abundance ratio in this region.

A more complete understanding of the likely existence of cool dust in the shock will require higher angular resolution and a broader wavelength coverage than that achieved by {\it Spitzer}.

Despite the faintness of emission from the main shock, the 24$\mu$m
map presents a new result, which was less obvious in previous studies,
namely that the dominant regions of star formation in SQ lie not in
the galaxies themselves, %(with the possible exception of NGC 7319) 
but
in two strikingly powerful, almost symmetrically disposed regions at
either end of the shock.  The region to the North is the well studied SQ-A, but the region to the South (which we refer to as
7318b-south) is also very powerful and both regions lie at the
ends of the shock, as defined by the \Ht distribution. This may not be a coincidence, and
we will discuss this further in Section 6.1.  

The $[$Ne\,{\sc ii}$]$12.81\micron, $[$Ne\,{\sc iii}$]$15.56\micron\ and 
$[$S\,{\sc iii}$]$33.48\micron\ fine-structure lines, as mentioned above, are
also tracers of star formation in SQ as these lines  are often associated with
\HII\ regions.
% and as noted above, the $[$Ne\,{\sc ii}$]$, $[$Ne\,{\sc iii}$]$ and $[$S\,{\sc iii}$]$ fine-structure lines indicate regions of star formation as they are associated with \HII\ regions.
The $[$Ne\,{\sc ii}$]$12.81\micron\ emission appears to follow {\it both} the
\Ht and the star formation regions (see Fig. \ref{fig:ne}a), %, showing the contours for $[$Ne\,{\sc ii}$]$ 
%and $[$Ne\,{\sc iii}$]$, 
appearing more
extended in the South than the corresponding \Ht emission and flaring out where star formation regions, especially 7318b-south,
are observed optically, and through PAH emission (see Fig. \ref{fig:dust}).
%Thus the $[$Ne\,{\sc ii}$]$ appears to follow {\it both} the
%\Ht and the star formation regions. 
%We see that the higher
%ionisation line of $[$Ne\,{\sc iii}$]$ is much weaker in the shock,
%but its peak corresponds to the peak in the \Ht S(3) line.
% We therefore see Ne emission associated with both star formation and shock excitation.

\section{Global Properties of the Molecular Hydrogen in SQ}\label{spec}

The warm molecular hydrogen in
Stephan's Quintet follows the X-ray distribution in the main shock and in the ``bridge'' structure. This might suggest
that the molecular hydrogen is excited directly by the X-ray heating. However, we will
show that the \Ht emission exceeds by at least a factor of 3 the X-ray
luminosity from the various shocked filaments, thus ruling out direct
X-ray excitation from the shock. 

\begin{figure*}[!pt]
\begin{center}
\includegraphics[width=17cm]{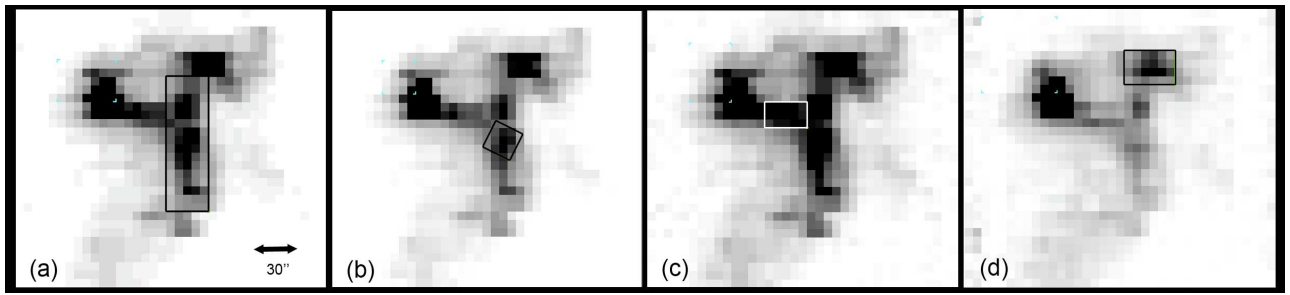}
\caption[]
{\small{The rectangular sub-regions extracted from the
main components of the SQ shock system overlaid on the \Ht 0-0
S(1) line emission. a) The Main Shock region centered on 22$^{\rm
h}$35$^{\rm m}$59.6$^{\rm s}$, +33$\degr$58\arcmin05.7\arcs. b)
A region in the main shock centered on 22$^{\rm h}$35$^{\rm
m}$59.97$^{\rm s}$, +33$\degr$58\arcmin23.3\arcs and chosen to
have limited contamination from star forming regions in SQ. c)
Extraction of the \Ht ``bridge'' feature centered on 22$^{\rm
h}$36$^{\rm m}$0.14$^{\rm s}$, +33$\degr$58\arcmin23.3\arcs. d)
An extraction of the SQ-A star forming region centered on 22$^{\rm
h}$35$^{\rm m}$58.85$^{\rm s}$, +33$\degr$58\arcmin50.4\arcs.}}
\label{fig:ext}
\end{center}
\end{figure*}

To measure the strength of the \Ht emission, we extract spectra from
various rectangular sub-areas of SQ which are defined in
Figure \ref{fig:ext}. The spectra were extracted from CUBISM cubes built
from each IRS module, and joined to make a continous spectrum--no
scaling was necessary to join the spectra.

\begin{figure*}[!p]
\begin{center}
\subfigure[Main Shock Region]{\includegraphics[width=8cm]{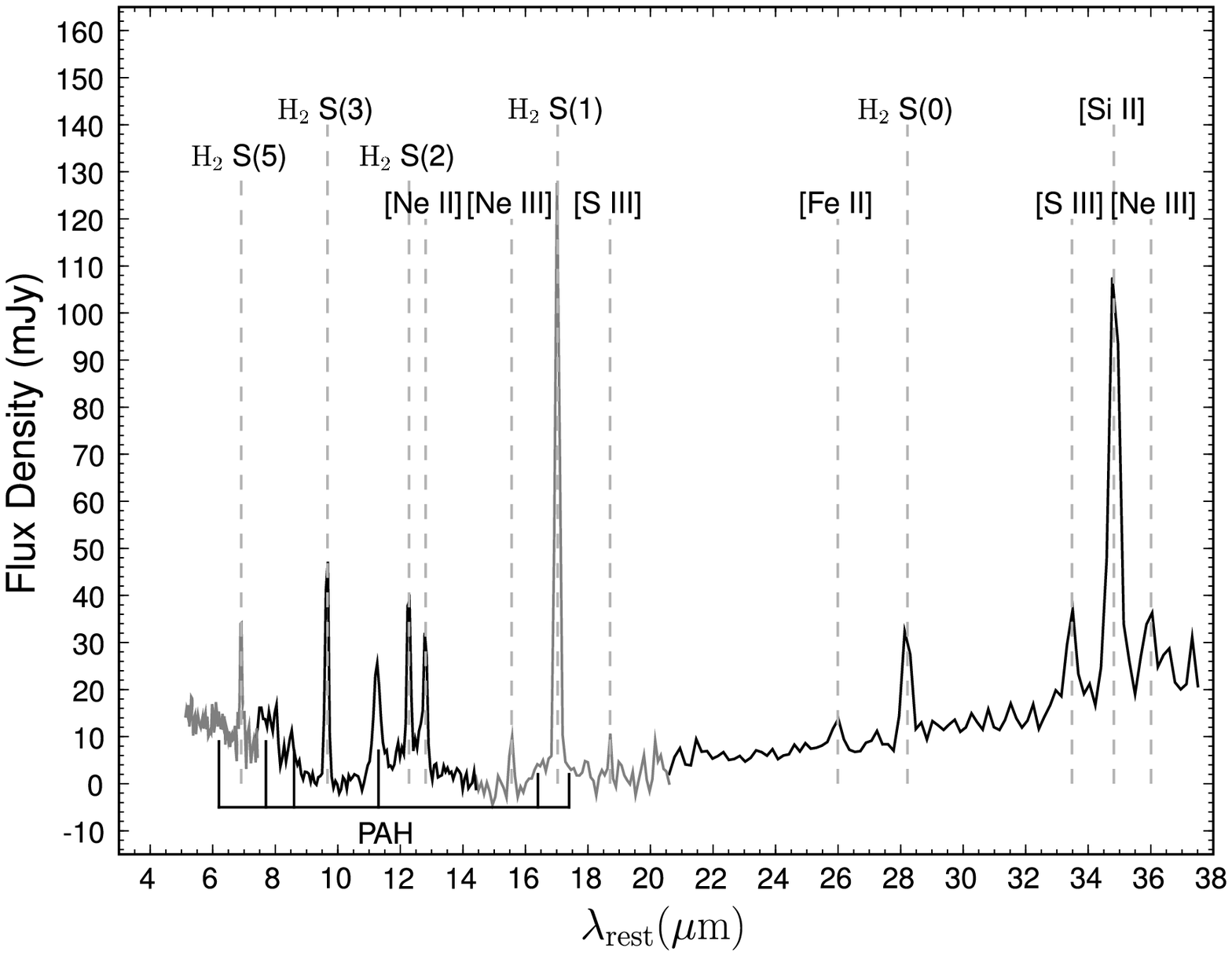}}
\hfill
\subfigure[Shock Sub-region ]{\includegraphics[width=8cm]{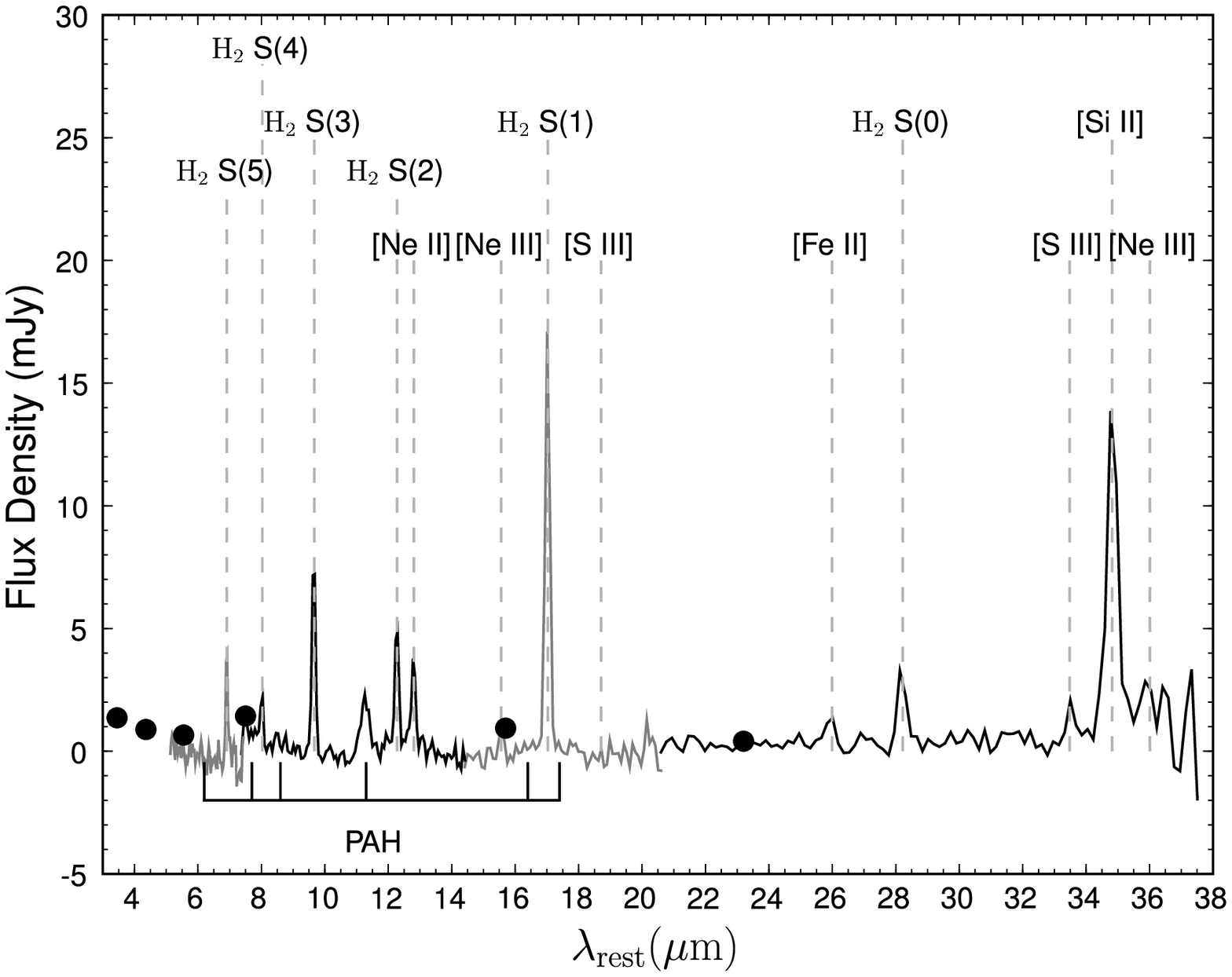}}
\vfill
\subfigure[``Bridge'']{\includegraphics[width=8cm]{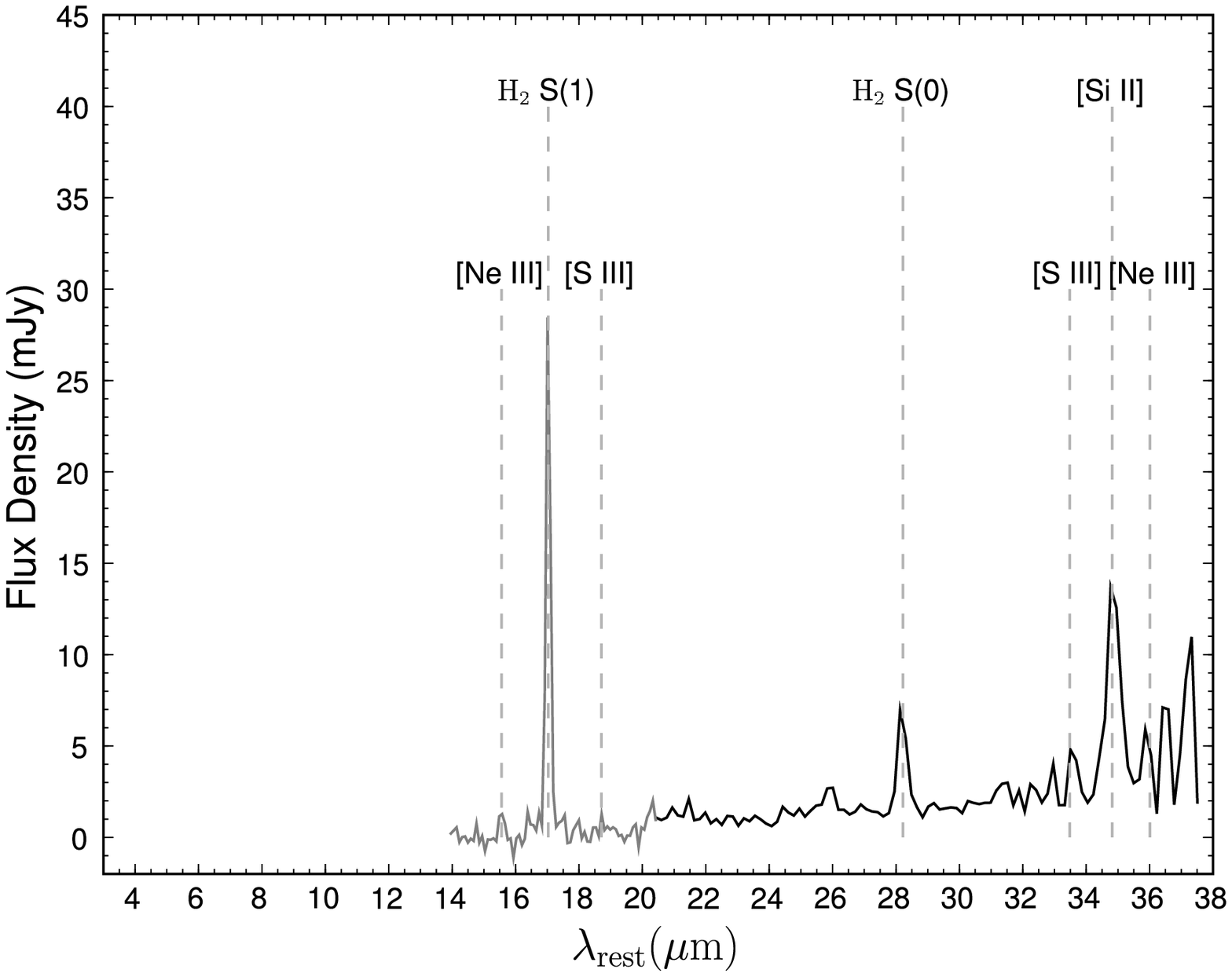}}
\hfill
\subfigure[SQ-A]{\includegraphics[width=8cm]{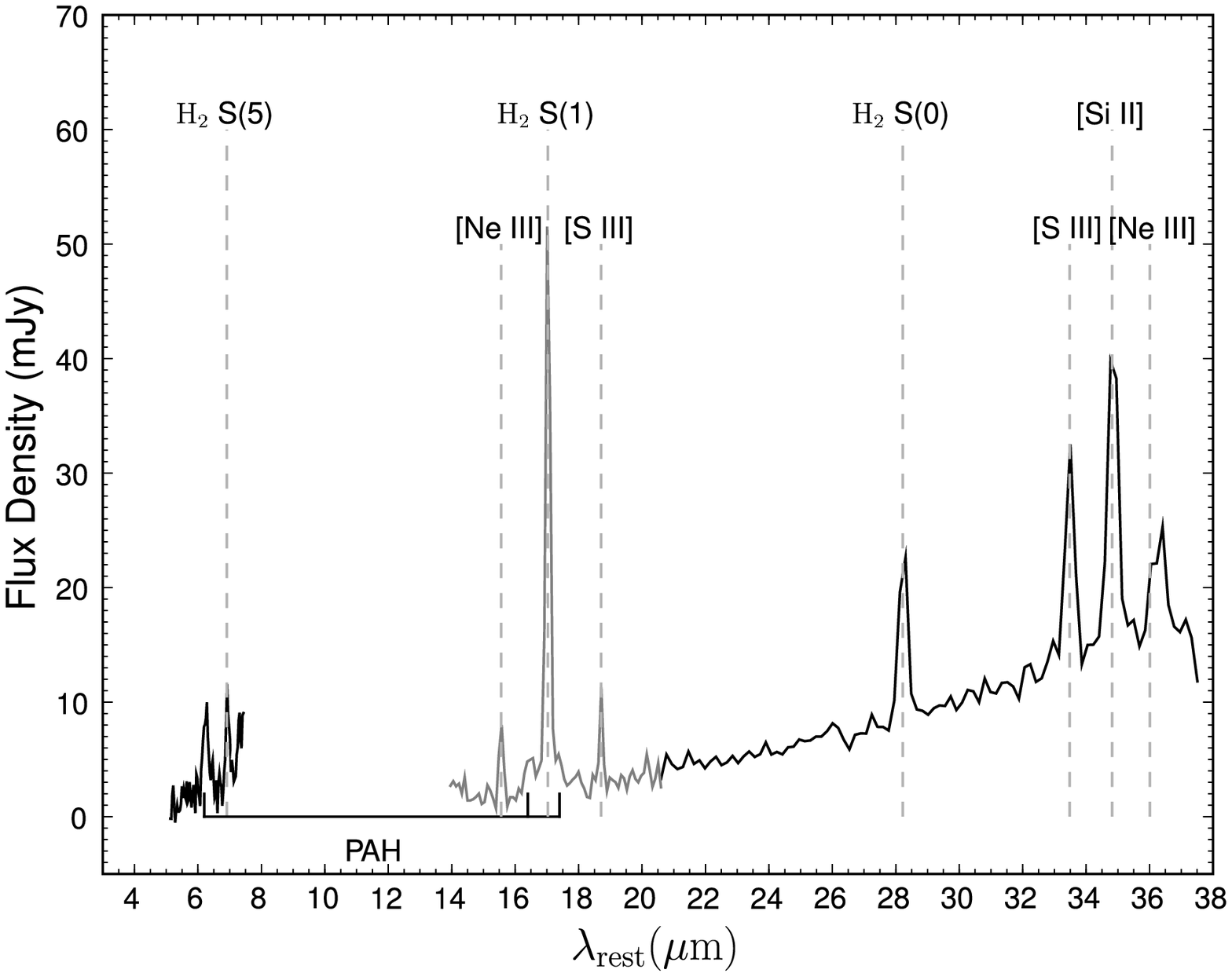}}

\caption[]
{\scriptsize{Extracted Spectra (low resolution) of the regions shown in Figure \ref{fig:ext}, combining all IRS modules with coverage. The key emission features are labeled with measurements and sizes of extraction areas given in Table~\ref{tableh2fluxes}, which provides the \Ht line fluxes, and Table~\ref{tablemetalfluxes} listing the metal line measurements. The shock sub-region (shown in b) includes the IRAC, 16\micron\ PUI and 24\micron\ MIPS measurements for the same area. The 16\micron\ flux is 0.939$\pm$0.047 mJy, which includes emission from the 17.03\micron\ \Ht line covered by the filter, while the 24\micron\ flux is 0.408$\pm$0.046 mJy; photometry was performed using a circular aperture with no aperture correction applied to compensate for contamination from the
surrounding emission. The corresponding IRAC fluxes are 1.362$\pm$0.068, 0.885$\pm$0.045, 0.659$\pm$0.036 and 1.434$\pm$0.073 mJy for the 3.6, 4.5, 5.8 and 8.0 \micron\ bands, respectively. The extraction of SQ-A (shown in d) also has coverage from the SL2 module due to the orientation of the IRS during the SL1 observation. }}
\label{fig:spectra}
\end{center}
\end{figure*}

Figures \ref{fig:ext}~a,~b and c indicate the spectral extraction regions of
the main NS shock, a sub-region of the shock and a
characteristic part of the ``bridge''. The shock sub-region is chosen
to be just North of the center of the shock, avoiding regions
contaminated by star formation in the intruder. Spectra for these
extractions are shown in Figure \ref{fig:spectra}~a, b and c. All
three spectra, share the common property that they are dominated by
molecular hydrogen emission. The shock sub-region
(Fig. \ref{fig:spectra}b), unlike Figure \ref{fig:spectra}a, is less
contaminated by the star forming regions discussed in the previous
section. The MIR continuum of the main shock appears stronger than the
shock sub-region indicating stronger emission from warm dust; this is
likely the result of contamination from star-forming regions in the
main shock extraction. The bridge exhibits a similarly weak continuum
emission compared to the shock sub-region.

Figure \ref{fig:spectra}b includes photometry from the IRAC bands, 16\micron\ Peak-up Image (PUI) and MIPS-24\micron\ image superimposed on the IRS spectrum. These are useful to probe conditions in the shock, in particular star formation (see Section 6.4). %The results strongly confirm the accuracy of the background subtraction and calibration of the IRS spectra. 
Stellar light in the extraction area, from an extended spiral arm of NGC 7318b, produces contamination of the shock spectrum, visible as continuum emission shortwards of 6\micron, in both Figure 8a and b.

There exists a striking similarity between the mid-IR spectrum of all three regions, showing powerful \Ht lines and low excitation
weak emission from fine structure lines. Also the PAH emission observed in the spectra of the shock regions, and in the region of the ``bridge'' in the IRAC 8\micron\ image (Fig. \ref{fig:dust}b), appears weak.
%The striking similarity between the mid-IR spectrum of all three
%regions (powerful \Ht lines, weak PAH emission, and low excitation
%weak emission from fine structure lines) 
This confirms that these
properties, observed in the \citet{App06} observations of the shock core,
extend to both the full extent of the main shock and the
``bridge''. This, and the fact that the ``bridge'' has similar X-ray
properties to the main shock (see later) suggests that the bridge
is a ``scaled-down'' version of the main shock. The weaker radio continuum emission at this location is significant. One possibility, that the ``bridge'' is older than the main shock, and the cosmic rays compressed in it have diffused away, will be discussed further in Paper II.Thus the new IRS
observations seem to suggest that more than one large-scale
group-wide shock is present in the group. This could be the result of previous tidal interactions and imply multiple shock heating events have taken place in SQ, consistent with what is seen in the X-ray \citep{Osul09}.

%{\bf discuss continuum differences - refer to Pierre's dust paper}

 Figures \ref{fig:ext}d and \ref{fig:spectra}d
present the extraction region and spectrum of SQ-A, the extragalactic star-forming
region. In this case, although \Ht lines are still strong, a rising continuum
and an increase in the strength of the metal lines relative to the \Ht is consistent with
a spectrum that is increasingly dominated by star formation -- a result which
is already known from previous optical observations \citep{Xu03}. 

Line fluxes for all the \Ht and metal lines in the spectra discussed above are presented in
Table~\ref{tableh2fluxes} and Table~\ref{tablemetalfluxes},
respectively. 
We estimate the luminosity emitted from the \Ht lines in the main
aperture shown in Figure \ref{fig:spectra}a. The emission from the 0-0
S(1) line alone can be calculated from Table~\ref{tableh2fluxes} (for
D = 94~Mpc) to be 2.3~$\times$10$^{41}$ erg\,s$^{-1}$. Summing the
emission measured in the observed lines for the main shock (0-0 S(0)
through S(5) lines) and including an extra 28\% emission from
unobserved lines (see model fit to excitation diagram below), yields a
total \Ht line luminosity from the main shock of
9.7~$\times$10$^{41}$ erg\,s$^{-1}$. This phenomenal power in the
molecular hydrogen lines dwarfs by a factor of ten the next brightest
mid-IR line, which is $[$Si\,{\sc ii}$]$34.82\micron\ with a line luminosity of
L$_{\rm SiII}$ = 0.85 $\times$10$^{41}$ erg\,s$^{-1}$.

\begin{figure*}[!tb]
\begin{center}
\includegraphics[width=12cm]{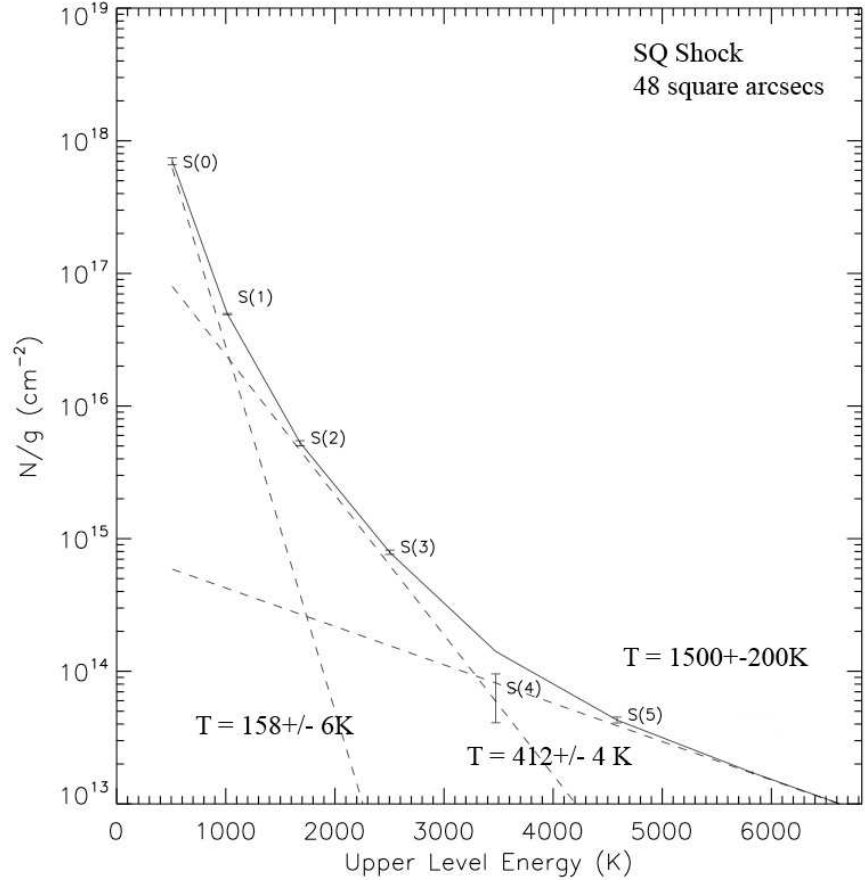}
\caption[]
{\small{Excitation Diagram of the Main Shock region in Stephan's Quintet. The ordinate (N/g) represents the column density divided by the statistical weight of the rotational state.}}
\label{fig:excite}
\end{center}
\end{figure*}

Figure \ref{fig:excite} presents the excitation diagram of the low-J
0-0 \Ht transitions for the main shock extraction. The points are well
fit by a model including three temperature components (T$_1$ = 158$\pm6$\,K,
T$_2$ = 412$\pm4$\,K, and T$_3$ = 1500$\pm200$\,K).  It is likely that in reality,
many different temperature components are present in the shock, and
the three-temperature fit is only an approximation. However,
it does allow us to provide an estimate of the total mass of warm \Ht
of 5.0$\pm 1 \times$10$^8$\,M$_\odot$. %{\bf Discuss cold from CO?}
Temperature T$_2$ is more uncertain than formally represented by the fit because it depends on the value of the S(4) flux, which may be systematically too low due to PAH contamination (see Fig. \ref{fig:excite}).
In Paper II we will present a more
complete two-dimensional map of the excitation of the \Ht in SQ and
explore variations in the shape of the excitation diagram along and
across the shock in more detail.

Our observations have shown that \Ht is the dominant line coolant in
the MIR. However, how does it compare with the most important
coolant in high-speed shocks--namely the X-ray emission? \citet{App06} suggested that the \Ht emission was stronger than
the X-ray emission at the shock center. This can now be evaluated over much of the inner SQ group.

We present a complete reanalysis of the XMM-Newton observations of SQ using the latest calibrations (see Appendix C for full details) in order to determine the fluxes and luminosities of the X-ray emission to match our spectral extractions. 
The results indicate the striking dominance of
the \Ht line luminosities compared with the X-ray emission from the
same regions. For the main shock, the X-ray ``Bolometric'' flux of
L$_{X(0.001-10 \rm keV)}$ = 2.8~$\times$~10$^{-13}$erg~s$^{-1}$cm$^{-2}$ corresponds to 2.95 $\times$~10$^{41}$~erg~s$^{-1}$, or
L(H$_{2}$)/L$_{X(0.001-10 \rm keV)}$ = 2.9. This is a lower limit since we
have not attempted to remove the contribution to the main shock
aperture of an extended group-wide X-ray component upon which the
emission from the shock lies. Therefore it is likely that the \Ht line luminosity dominates over the main shock X-ray gas by a factor $\ge$ 3. Similar calculations can be done for the
other regions for which \Ht spectra have been extracted. For example,
in the ``bridge'' region, which we have already indicated has many of
the same characteristics as the main shock, we find
L(H$_{2}$)/L$_{X(0.001-10 \rm keV)}$~=~2.5. These values
demonstrate that throughout the extended regions of SQ, the
molecular hydrogen cooling pathway dominates over the X-ray in this shocked system. %contribution in these high-speed shocks is molecular
%hydrogen. 
This is a very significant result, upturning the
traditional view that X-ray emission always dominates cooling in the
later stages of evolution in compact groups of galaxies.

%We will return to this theme later in the paper.

%{\BF------------------------end of new section on gobal h2 properties}

\section{Emission Line Diagnostics}\label{em}

The fine-structure flux ratios (see Table \ref{tablemetalfluxes}) can
be used to probe the conditions within the extracted regions of
SQ. The $[$Si\,{\sc ii}$]$\,34.82\micron/$[$S\,{\sc iii}$]$33.48\micron\ ratio provides an
indication of the sources of excitation within the system.  As
mentioned previously, $[$S\,{\sc iii}$]$ is mainly a tracer of \HII
regions, whereas enhanced $[$Si\,{\sc ii}$]$ emission can be generated
via several mechanisms, including thermal excitation by X-rays (XDRs),
or in shocks.  In the main shock, the $[$Si\,{\sc ii}$]$\,34.82\micron/$[$S\,{\sc iii}$]$33.48\micron\ ratio of $\sim$ 4.59 is high compared to, for example, both normal galaxies ($\sim$ 1.2) and AGN ($\sim$ 2.9)
in the SINGS sample \citep{Dal06}. However, this large aperture is contaminated by star
formation emission from SQ-A and the intruder galaxy. A better measure is given by the smaller
shock sub-region, where the $[$Si\,{\sc ii}$]$\,34.82\micron/$[$S\,{\sc iii}$]$33.48\micron\ ratio is $\sim9.5$. Thus it is clear that the $[$Si\,{\sc ii}$]$ emission is well
outside the normal range of values, even for local well-studied
AGN. Using the upper limit found for $[$S\,{\sc iii}$]$\,33.48\micron\ 
in the ``bridge'' structure, we find a
ratio of $>5$ -- again values well outside the range of normal galaxy
disk emission.  Indeed, these high values are typical of galactic
supernova remnants where shock excitation is well determined
\citep[e.g.][]{Neu07,Hew09}.  We
will argue in the next section that silicon is being ionised in regions experiencing fast shocks $100 < V_{s} < 300$\,km\,$\rm{s^{-1}}$ and depleted onto dust grains.
The $[$Fe\,{\sc ii}$]$\,26.0\micron/$[$Si\,{\sc ii}$]$\,34.82\micron\ ratio for the main shock and sub-region ($\sim$ 0.12) is in agreement with values found by \citet{Neu07} for their sample of SNR.

For the extragalactic star forming region SQ-A, $[$Si\,{\sc ii}$]$\,34.82\micron/$[$S\,{\sc iii}$]$\,33.48\micron\ is $\sim$1.49, only slightly higher than the
average of $\sim$~1.2 found for star-forming regions in the SINGS sample
\citep{Dal06}; another indication that SQ-A is dominated by star formation.

%Metallicity (\citet{Dal09}: 

%Main shock: 12 + log(O/H) = 8.9

%The $[$Ne\,{\sc ii}$]$\,12.81\micron\ and $[$Ne\,{\sc iii}$]$\,15.56\micron\ luminosities can be used as measures of star formation \citep{Des07,Ho07}. From this we obtain a SFR of 3.7\,$M_{\odot}\, \rm{yr^{-1}}$ in the main shock. This relatively high value is likely due to contamination from star forming regions in the intruder galaxy and does not imply a high SFR in the shock itself. This is discussed quantitatively in section \ref{SFS}.

In star-forming galaxies the $[$Ne\,{\sc iii}$]$\,15.56\micron/$[$Ne\,{\sc ii}$]$\,12.81\micron\ ratio can be used as a measure of the hardness of the radiation field as it is sensitive to
the effective temperature of the ionising sources. In the main shock,
we find a value of $\sim0.36$ which would be considered typical compared to those found in starburst systems,
which range from $\sim0.03-15$ \citep{Ver03,Ber09}, and in supernovae remnants ranging from $\sim0.07 - 1.24$ \citep[][]{Neu07,Hew09}. The shock
subregion has a ratio of only $\sim0.14$ indicating a lower intensity radiation field North of the shock center. However, it is clear from its spatial distribution relative to the 8\micron\ and \Ht emission, that most of the neon is not originating from star formation, and so shocks are an obvious source of excitation.

The $[$Ne\,{\sc ii}$]$\,12.81\micron/$[$Ne\,{\sc iii}$]$\,15.56\micron\ ratio can be used to estimate the shock velocity using the MAPPINGS shock model library of \citet{All08}. In the main shock this %we find a ratio of $\sim$~2.74  which corresponds %\citep[from the models of][]{Hart87}, 
ratio corresponds to shock speeds of between 100 and 300\,km\,s$^{-1}$ (using preshock densities of $0.01< n < 100$ cm$^{-3}$ and magnetic parameter $B/n^{0.5}$ = 1 and 3.23 -- the nominal equipartition value\footnote{See Allen et al.(2008)}).
We do, however, have contamination from starforming regions in the main shock and cannot disentangle this emission from that produced in the shock. To address this we use the $[$Ne\,{\sc ii}$]$\,12.81\micron\ and $[$Ne\,{\sc iii}$]$\,15.56\micron\ maps to mask areas associated with star formation (see Section 3.3) and determine a lower limit for the $[$Ne\,{\sc ii}$]$\,12.81\micron/$[$Ne\,{\sc iii}$]$\,15.56\micron\ ratio in the shock of $\sim$~4.54. This value corresponds to a shock velocity of $\sim$ 150\,km\,s$^{-1}$.
%We shall use this in Section \ref{fe_si} to explore depletion onto dust in the shock.

%200\,km\,s$^{-1}$, 100\,km\,s$^{-1}$ and/or 80\,km\,s$^{-1}$ and the shock sub-region value corresponds to a shock of 80\,km\,s$^{-1}$. 
%we find that the $[$Ne\,{\sc iii}$]$\,15.56\micron/$[$Ne\,{\sc ii}$]$\,12.81\micron\ ratio for the main shock corresponds to shock velocities 

The average
electron density is determined from the
$[$S\,{\sc iii}$]$18.71\micron/$[$S\,{\sc iii}$]$33.48\micron\ (two lines of the same ionisation
state) ratio. For SQ-A we find a ratio of $\sim$0.56, and in the main shock $\sim$0.41. This corresponds to an electron density of
$100-200$ cm$^{-3}$ \citep{Mar02} for both, i.e. in the low-density limit
for this diagnostic \citep{Smi09}.

%Add discussion of IRS spectrum of this galaxy including the P200''
%spectrum from TripleSpec obtained last year. Is NGC 7319 a MOHEG?
%Answer this from line ratio and strength diagnostics. (Table 1 and 2).

\section{The Group-wide Shocks}\label{glob_prop}

\subsection{Origin of the H$_{2}$ and X-ray emission}

Our observations have shown that the molecular hydrogen and X-ray emitting plasma appear to follow a similar distribution, and we have ruled out the possibility that this is a 
consequence of X-rays heating the H$_{2}$, since the \Ht has the dominant luminosity.
How then can we explain the similar distributions? Are these results consistent with
the hypothesis that the shock is formed where the intruding galaxy NGC 7318b collides
with a pre-existing tidal filament of \HI drawn out of NGC 7319 in a previous interaction
with another group member \citep{Mol97, Trin03}?

\begin{figure*}[!tb]
\begin{center}
\includegraphics[width=15cm]{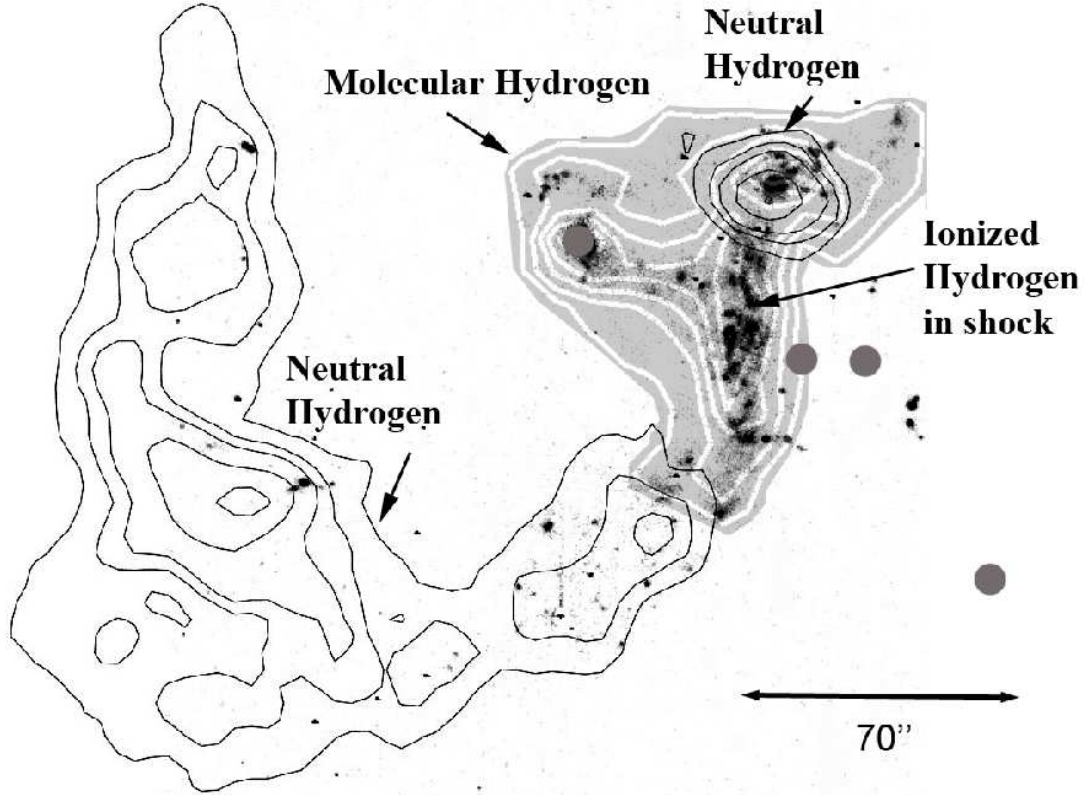}
\caption[]
{\small{Schematic Diagram of the \HI\ distribution in Stephan's Quintet, from \citet{Will02}, and the H$_{\alpha}$ emission \citep{Xu99}, with the \Ht 0-0 S(1) emission overlaid. The grey dots represent the locations of the core galaxies in the group as shown in Fig. \ref{fig:int}.}}
\label{fig:HI+H2}
\end{center}
\end{figure*}

This basic mechanical picture appears plausible as can be seen
in Figure \ref{fig:HI+H2} which shows that the \Ht distribution  ``fills in''
the gap in the \HI tidal tail as observed by the VLA \citep{Will02}. 
The implication is that the \HI has been
converted into both a hot X-ray component and a warm \Ht component by
the collision of the intruder with the now missing H\,{\sc i}.

Part of the puzzle of how this high-speed ($V_{\rm s}$ $\sim 700-1000$ km\,s$^{-1}$) shock
can lead to both X-ray and very strong molecular line emission is
presented in a model by \citet{Pg09}. The high-speed collision of
NGC~7318b with the \HI filament (assumed to be composed of a
multiphase medium) leads to multiple shocks passing through and
compressing denser clumps (which become dusty nucleation sites for \Ht
formation) as opposed to the lower-density gas, which is
shock-heated to X-ray tempertures. The \Ht
therefore forms in denser clouds experiencing slower shocks. Thus
the coexistence of both hot X-ray gas, and cooler molecular material 
is a natural consequence of the multiphase medium of the pre-shocked 
material. 

Modeling of the \Ht excitation by
\citet{Pg09} demonstrates that the emission can be reproduced by low velocity
($\sim 5-20\, {\rm km\, s^{-1}}$) magnetohydrodynamic shocks within
the dense ($n_{\rm H} > 10^{3}\, {\rm cm^{-3}}$) \Ht gas. The denser clouds survive
long enough to be heated by turbulence in the hot-gas component,
tapping into the large available kinetic energy of the shock. This
picture is consistent with both the broad \Ht linewidth (870\,km\,s$^{-1}$) measured in IRS high-resolution spectrometer observations of \citet{App06}, and the velocity center of the warm \Ht (based on new IRS spectral calibrations - see Appendix B) which places the gas at intermediate velocities between the intruder and the group IGM. Both these measurements are consistent with \Ht being accelerated in a turbulent post-shocked layer.
% ; unfortunately our study lacks the spectral resolution to probe the \Ht line width. However, the improved calibration of the high resolution data confirms that the \Ht velocity lies intermediate to the intruder and group (see Appendix B), consistent with dense clouds being accelerated in the shock.

Intermediate pre-shock densities and post-shock temperatures result in
regions of \HI and \HII that have cooled, but where the dust content
has been destroyed \citep{Pg09}. Pre-existing Giant Molecular Clouds (GMCs) which may have been embedded in the \HI\ gas, would be rapidly compressed and collapse quickly, thus forming stars.%by the shock, but not destroyed, causing instabilities that culminate in a burst of star formation. 
This mechanism, proposed for SQ-A by \citet{Xu03}, might also apply to 7318b-south. However, if this was the case, it would have to explain why two such GMCs happened to be positioned at the extreme ends of the current shock -- an unlikely coincidence. More probable, however, is that the geometry of the shock somehow favors the collapse of clouds at the ends of the shock -- perhaps in regions where the turbulent heating is less efficient.
%it seems unlikely for these GMCs to coincidentally lie at either end of the shock ridge. The star formation in SQ-A has two components, in the IGM as well as in the ISM of the intruder \citep{Xu03}, whereas 7318b-south appears to be associated with the intruder only \citep{Xu05}. It may be that due to the geometry of the shock, these regions are dynamically more favourable for star formation as the postshock medium here is less turbulent and molecular gas has had time to cool and collapse. Detailed kinematic modeling of the collision may be possible with future improved resolution data and remains to be explored.

%**show figure that puts the H2 emission
%into the neatral hydrogen map of the large-scale HI filaments and how
%it ``fills the gap'' caused by the X-ray shock**

\subsection{Origin of $[$Fe\,{\sc ii}$]$ and $[$Si\,{\sc ii}$]$ emission}\label{fe_si}

As outlined above, the combination of emission detected in the shock
region of SQ can be understood in terms of a spectrum of shock
velocities. The fastest shock velocities ($V_{s} \sim 600$
km\,$\rm{s^{-1}}$) are associated with the lowest density pre-shock
regions and the post-shock X-ray emitting plasma. These are fast $J$
shocks and represent a discontinous change of hydrodynamic variables
and are often dissociative \citep{Hol89}. $C$ shocks have a broad
transition region such that the transition from pre-shock to
post-shock is continuous and are usually non-dissociative
\citep{Dr83}. The lowest velocity shocks associated with the turbulent
\Ht emission are $\sim 5-20\, {\rm km\, s^{-1}}$ $C$ shocks
\citep{Pg09}. \citet{Xu03} find optical emission line ratios consistent with shock models that do not include a radiative precursor \citep{Dop95}.

\begin{figure*}[!t]
\begin{center}
\includegraphics[width=15cm]{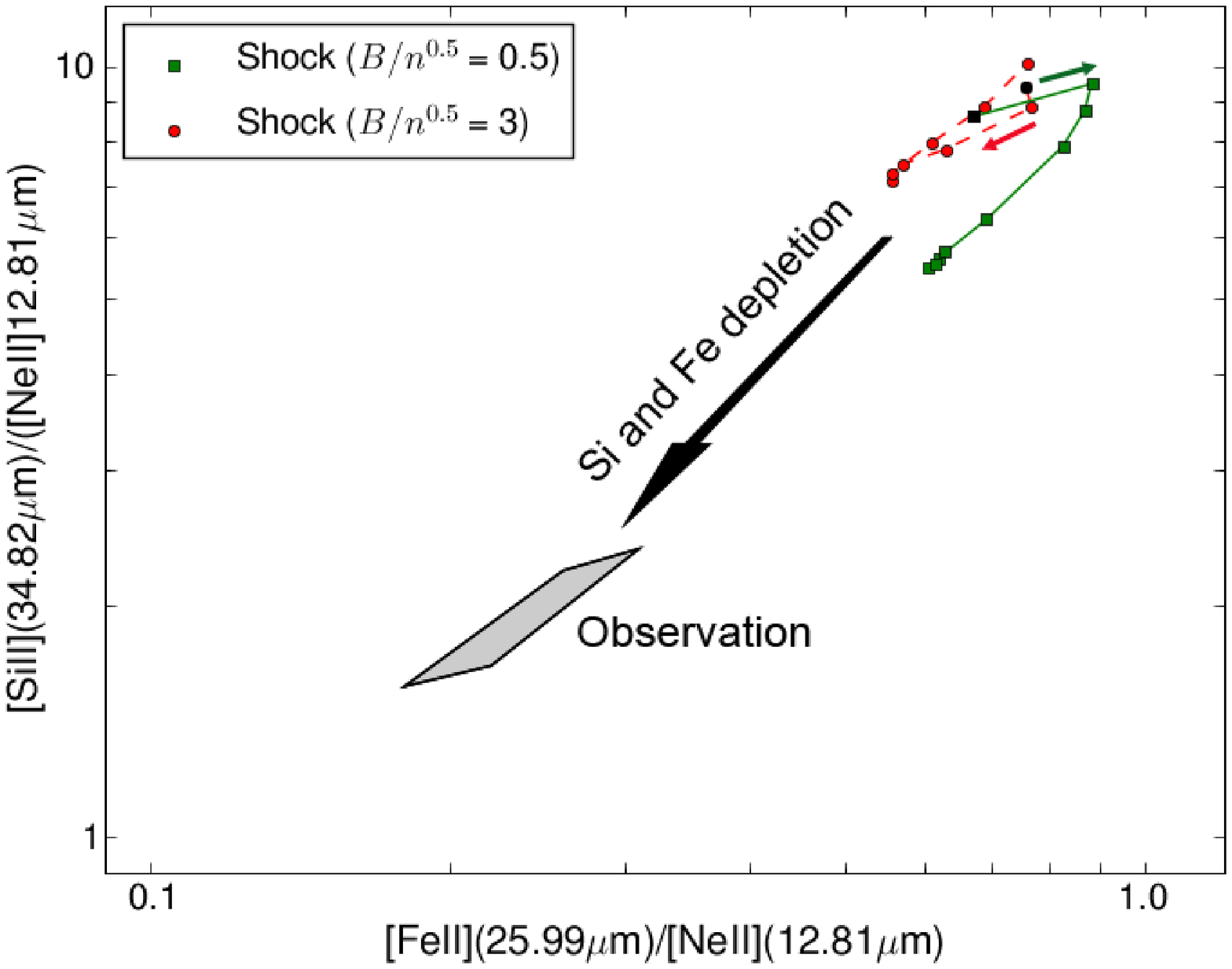}
\caption[]
{\small{The line ratios measured in the Main 
SQ shock of the 
$[$Ne\,{\sc ii}$]$($12.8\mu$m), $[$Fe\,{\sc ii}$]$($26.0 \mu$m), and 
$[$Si\,{\sc ii}$]$($34.8 \mu$m) lines are compared with 
shock calculations from Allen et al. (2008).
The model values cover a shock velocity range of
$100-300~$\,km\,s$^{-1}$ with a gas density of
$1\,$cm$^{-3}$ and two values of the magnetic
parameter $B/n^{0.5}$ = 0.5 (green square) and 3 (red circles). The black symbols represent the 100\,km\,s$^{-1}$ values and the arrows indicate the direction of increasing velocity (in steps of 25\,km\,s$^{-1}$). The grey area indicates the range of line
ratios, $[$Si\,{\sc ii}$]$/$[$Ne\,{\sc ii}$]$ and $[$Fe\,{\sc ii}$]$/$[$Ne\,{\sc ii}$]$, determined from observations (as discussed in Section 5). The ratios are lower than the model values which we interpret as evidence for Si and Fe depletion onto dust grains. }}
\label{fig:sil_ne}
\end{center}
\end{figure*}

%The energetic requirements for shocks to produce strong $[$Fe\,{\sc ii}$]$ 
%emission are usually present in $J$ shocks while the ion
%abundance in $C$ shocks are low in comparison \citep{Hol89}. 
The $[$Fe\,{\sc ii}$]$25.99\micron\ emission associated with the shock region
in SQ is relatively weak, but coincides with the most energetic part of the shock
as traced by the X-rays (Fig. \ref{fig:fs}). We also detect abundant $[$Si\,{\sc ii}$]$34.82\micron\ emission associated with the main shock.

Silicon and iron have very similar first and second ionization
potentials. Their first ionization potentials (7.9 and 
8.15~eV for Fe and Si, respectively) are lower than 
that of hydrogen but their second ionization potential 
is higher (16.19 and 16.35~eV). 
The MIR $[$Fe\,{\sc ii}$]$ and $[$Si\,{\sc ii}$]$ line emission observed from the SQ shock 
could thus arise from predominantly neutral, as well as ionized gas.
We discuss the contribution from the ionized gas using the 
$[$Fe\,{\sc ii}$]$($25.99 \mu$m)/$[$Ne\,{\sc ii}$]$($12.81\mu$m) and $[$Si\,{\sc ii}$]$($34.82
\mu$m)/$[$Ne\,{\sc ii}$]$($12.81\mu$m) line ratios. 
The high $[$Ne\,{\sc ii}$]$($12.81\mu$m)/$[$Ne\,{\sc iii}$]$($15.56\mu$m) mid-IR line ratio (see Section 5) implies that
$[$Ne\,{\sc ii}$]$ is the dominant ionization state of Ne in the SQ shock. 
Unlike Fe and Si, Ne is not much depleted on dust \citep{Sim98}. Because the $[$Ne\,{\sc ii}$]$\,$12.81\mu$m line
has a high critical density ($n_e = 4.3 \times 10^5~$cm$^{-3}$,
Ho and Keto 2007), the neon line strength scales with the emission measure of the ionised gas.

The optical line emission from the SQ shock is discussed in detail in
Xu et al. (2003). The high values of the $[$O\,{\sc i}$]$(6300\AA) and
$[$N\,{\sc ii}$]$(6584\AA) lines to H$\alpha$ line ratios are evidence of 
shock ionization. The optical $[$S\,{\sc ii}$]$(6716\AA/6731\AA) line ratio as
well as the MIR $[$S\,{\sc iii}$]$ line ratio correspond to the low density
limit (see Section 5) and comparison with the shock models of Allen et al. (2008) 
constrain the pre-shock gas density to be about $1~$cm$^{-3}$ or
smaller. 

In Figure \ref{fig:sil_ne} we indicate the region (in grey) corresponding to the observed MIR ratios (using the upper and lower limits determined for the $[$Ne\,{\sc ii}$]12.81\mu$m line emission as discussed in Section 5). For comparison the expected emission from the shock models of \citet{All08} are displayed for 
%shock velocities from 100 to 300\,km\,s$^{-1}$ with 
a pre-shock gas density of $1~$cm$^{-3}$ and two values of the magnetic parameter, $B/n^{0.5}$ = 0.5 and 3.
% The IRS observations are compared to shock models from 
%Allen et al. (2008) in Figure \ref{fig:sil_ne}. 
%Model values for both the shock and the shock plus its radiative 
%precursor  are  displayed for 
%a pre-shock gas density of $1~$cm$^{-3}$ and for two values
%of the magnetic parameter, $B/n^{0.5}$ = 0.5 and 3.
%The shock velocity ranges from 100 to 300\,km\,s$^{-1}$ in steps
%of 25\,km\,s$^{-1}$. 
For clarity of the figure, 
only shock velocities from 100 to 300\,km\,s$^{-1}$ are used and higher velocities shocks which 
do not match the observed $[$Ne\,{\sc ii}$]$/$[$Ne\,{\sc iii}$]$ line emission ratio (see Section 5) are discarded.

%Note that the visible emission line ratios 
%put the SQ shock within the LINERs classification
%of optical spectra which is known to be best fit by shock
%models without emission from the radiative precursor \citep{Dop95}.
%The absence of the precursor component could be accounted for by
%the clumpiness of the interstellar medium which may make it 
%optically thin to ionizing radiation. Most of the gas mass
%is observed to be in H$_2$ gas at high densities (n$_{\rm H}> 10^3$cm$^{-3}$) 
%that must have a small volume filling factor (Guillard et
%al. 2009). 

The observed $[$Fe\,{\sc ii}$]$($25.99 \mu$m)/$[$Ne\,{\sc ii}$]$($12.81\mu$m) and $[$Si\,{\sc ii}$]$($34.82
\mu$m)/$[$Ne\,{\sc ii}$]$($12.81\mu$m) ratios are both smaller than the shock values.  
The iron and silicon lines are not dominant cooling lines 
of ionizing shocks. The gas abundances of these elements do not
impact the thermal structure of the shock and the 
line intensities roughly scale with the gas phase abundances.
We thus interpret the offset between the IRS observation and 
model values as evidence for Fe and Si depletion.  
We consider the magnitude of the depletions indicated by the arrow ($\sim$50\% and $\sim$60\% for Fe and Si, respectively) in Figure \ref{fig:sil_ne} 
as lower limits since there could be a contribution 
to the $[$Fe\,{\sc ii}$]$ and $[$Si\,{\sc ii}$]$ line emission from non-ionizing $J$-shocks into 
molecular gas \citep{Hol89}. 
Such shocks could also contribute to the $[$O\,{\sc i}$]$(6300\AA) line emission as discussed
in Guillard et al. (2009).
Forthcoming observations of the H$_2$ ro-vibrational line emission 
in the near-IR should allow us to estimate 
whether they may be significant.
A contribution
from  non-ionizing shocks to the $[$Fe\,{\sc ii}$]$ and $[$Si\,{\sc ii}$]$ line emission
will raise the depletion of both Fe and 
Si as well as the Fe/Si depletion ratio because  non-ionizing shocks do
not produce $[$Ne\,{\sc ii}$]$ line emission and  
the $[$Fe\,{\sc ii}$]$($25.99\mu$m)/$[$Si\,{\sc ii}$]$($34.82 \mu$m)
line emission ratio in $J$-shocks within dense gas is larger than for 
shocks plotted in Figure \ref{fig:sil_ne}. 

\subsection{Comparison with Cold H$_{2}$ Distribution}\label{CO}

\begin{figure*}[!t]
\begin{center}
\includegraphics[width=17cm]{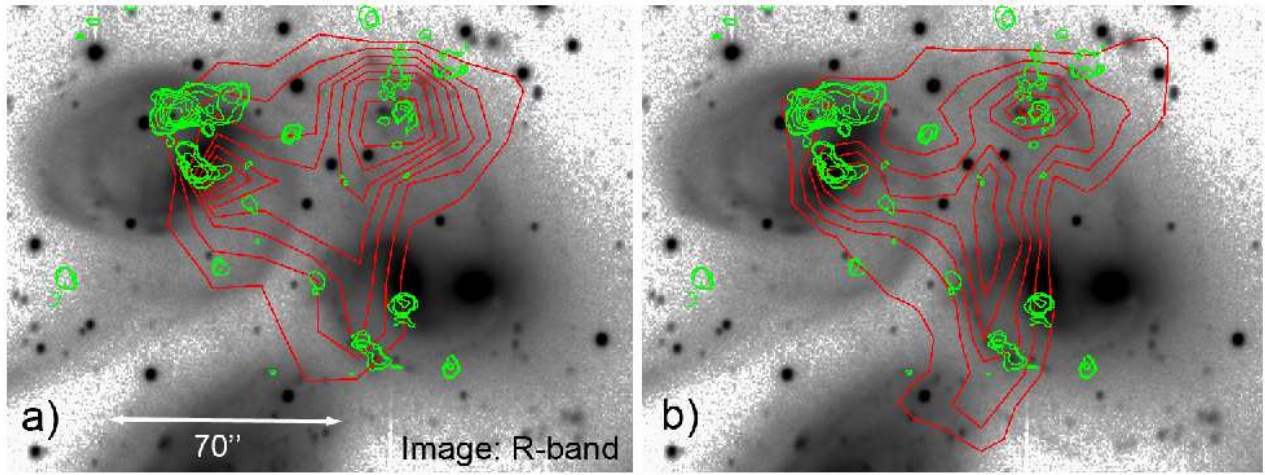}
\caption[]
{\small{CO (1-0) contours (green) from \citet{Gao00} overliad on an R-band image \citep{Xu99} of the group. Contour levels correspond to 2.4$\sigma$, 3.0$\sigma$, 4.0$\sigma$, 5.0$\sigma$, 6.0$\sigma$, 7.0$\sigma$ and 8$\sigma$ (1.0$\sigma \sim$0.025 Jy). We show the H$_2$ contours (red) for the S(0) line (a) and the S(1) line (b) with contour levels as in Fig. \ref{fig:opt_h}a and b respectively.
}}
\label{fig:CO}
\end{center}
\end{figure*}

A key aspect to understanding the emission we observe in SQ, as well as a test of the proposed model of \Ht excitation, is the amount and distribution of cold H$_{2}$, as measured by CO. A reservoir of cold \Ht associated with the warm gas observed in SQ would provide key insight into conditions within the intragroup medium.
In Figure \ref{fig:CO} we show the BIMA CO (1-0) integrated intensity contours from \citet{Gao00}, shown with the \Ht S(0) and S(1) emission contours, overlaid on an optical image of SQ. These interferometric observations use the large primary beam of 110\arcsec, centered on the shock, to determine areas of high column densities in the group. The CO traces areas of known star formation seen at 24\micron\ and in the NUV (see Section \ref{SF}), notably SQ-A and 7318b-south. As observed by \citet{Yun97}, the CO around NGC 7319 is concentrated in two regions. The dominant complex is North of the nucleus residing in a dusty tidal feature. The nuclear CO is elongated perpendicular to the stellar disk suggesting a deficiency of ongoing star formation in the disk.

The CO distribution does not correlate with the location of warm \Ht emission, particularly around NGC 7319. The cold \Ht complexes are clearly offset from the concentrations of warm H$_2$. Even the S(0) emission line (see Fig. \ref{fig:CO}a), which follows the coldest warm H$_2$, does not have peaks corresponding to the strongest CO detections.%, with cold gas concentrations appearing to residing furthrest from the shock by comparison. We find no concentrations of CO associated with the main shock or bridge structure where we find copious warm emission. 
 
%\citet{Lis02} mapped CO in
%several regions of the IGM of SQ using the IRAM 30 m antenna, focusing
%mainly on the starbursts SQ-A and SQ-B. They did, however, detect narrow-line width CO
%emission above 3$\sigma$ (T$_{\rm mb} \sim$9 mK) along the N-S shock ridge (as defined by th 20 cm and
%X-ray maps) North of NGC 7318b, with most of the gas at a velocity of
%6000 km\,s$^{-1}$, but also detected at 6700 km\,s$^{-1}$. 

In a forthcoming paper (Guillard et al. 2010, in preparation), we will report the recent detection of $^{12}$CO(1-0)
and (2-1) emission, associated with the warm
H$_2$ in the SQ shock, using the single-dish 30m IRAM telescope. These observations suggest that most of the CO
emission in the shock has been missed by interferometers (because of the broad linewidth) and
show that the CO emission is both present in the shock, and extends along the H$_2$-emitting
bridge and towards NGC 7319. % (in the so-called H$_2$-bridge). 

The kinematics of the CO gas lying
outside of star-forming regions, in the new observations, appears to be highly disturbed with a broad linewidth in agreement with the Appleton et al. (2006) interpretation that the MIR \Ht lines were intrinsically very broad and resolved 
by the high resolution module of IRS. The CO data also agrees with our re-analysis of the Appleton et al. (2006) data (see Appendix B)  using more recent and reliable wavelength calibration, which places 
the bulk of the \Ht gas at velocities intermediate to that of the intruder and the group -- supporting the idea that the \Ht gas is accelerated in the shock.

%The
%kinematics of the CO gas lying outside of star-forming regions appears to be
%highly disturbed;this agrees with the reanalysed high resolution spectrum (shown in Appendix B) that the molecular hydrogen emission is intrinsically broad and the gas is being accelerated in the shock

%Deep IRAM
%CO observations in the shocked emission regions of SQ have detected substantial $^{12}$CO(1-0) broad line emission at a velocity between the intruder and the group (P. Guillard -- Private Communication). This agrees with the reanalysed high resolution spectrum (shown in Appendix B) that the molecular hydrogen emission is intrinsically broad and the gas is being accelerated in the shock. However, detailed modeling is beyond the scope of this paper and will be discussed in Guillard et al. (2009, in preparation).

%These deep
%observations suggest that the warm \Ht emission is the main molecular component
%in the central part of the shock-- consistent with a model of turbulent heating discussed earlier. 

\subsection{Turbulent Suppression of Star Formation in the H$_{2}$ filaments}\label{SFS}

Section \ref{SF} discussed the star forming regions observed in SQ
and noted that there was very little evidence for star formation in
the shock associated with the warm \Ht emission. The main shock
contains a large quantity of warm molecular hydrogen (5.0
$\times$10$^8$\,M$_\odot$) providing a reservoir of fuel for star
formation once it cools (see Section \ref{spec}).
We investigate star formation in the warm H$_2$-dominated medium by considering the shock sub-region, chosen to avoid star forming regions in the intruder galaxy, but likely still subject to some contamination from these regions.

PAH emission is a classical tracer of star
formation, but the molecules are fragile and easily destroyed in hard radiation fields. In the spectra of the main shock and subregion (Fig. \ref{fig:spectra}a and b) the PAH emission bands at 6.2
\micron, 7.7 \micron\ and 8.6 \micron\ are far weaker compared to the
11.3 \micron\ bands, which when strong are predominantly produced by
neutral PAH molecules \citep{Dr01}.
In the shock subregion we find an upper limit flux for the 6.2 \micron\ PAH of $0.47 \times 10^{-17}$ W\,m$^{-2}$ and fluxes of $0.87 \times 10^{-17}$ W\,m$^{-2}$ and $2.49 \times 10^{-17}$ W\,m$^{-2}$ for the 7.7 and 11.3 \micron\ bands respectively. This corresponds to a 7.7\micron/11.3\micron\ PAH ratio of 0.35, very low compared to the median value found for the SINGS sample of 3.6 \citep{Smi07}. 

The suppression of the 7.7\micron/11.3\micron\ PAH ratio has been observed in AGN environments \citep[e.g. low-luminosity AGN in the SINGS sample of][]{Smi07} and is favored to be the result of selective destruction of PAH molecules small enough to emit at 7.7\micron. PAH processing in the shock due to larger molecules being less fragile than smaller ones is discussed further in Guillard et al. (2010). 
An alternative explanation is that the PAH molecules are chiefly large and neutral in the shock, producing enhanced 11.3\micron\ emission in comparison to the smaller PAH molecules. %This could be the result of a modified distribution of molecular carriers favoring the 11.3\micron\ band. 
A detailed comparison of the dust and PAH emission properties can be found in Guillard et al. (2010).

%This implies that the PAH environment modifies the PAH grain size distribution as the different ionisation states are affected differently. This could be due to differing grain or molecular carriers for these two bands. The PAH dust properties and its implications for the ionisation state in the shock are .

%The lack of MIR continuum is
%striking, implying very little emission from VSGs (Very Small Grains)
%in the shock.

%\begin{figure*}[!th]
%\begin{center}
%\includegraphics[width=8.7cm]{shockplot_noSF_all_bw.eps}
%\caption[]
%{\small{Region in shock not contaminated by star formation in the intruder}}
%\label{fig:noSF}
%\end{center}
%\end{figure*}

%{\bf PAH suppression}

%Despite this region being chosen to be representative of the star
%forming conditions occurring in the postshock medium (i.e. avoiding
%star forming regions in the intruder galaxy) it is still subject to
%some contamination from these regions. 
Dust emission in the mid- and far-infrared can be used to infer the amount of star formation taking
place \citep{Ken98}. We measure a 24\micron\ flux in this region of
0.408 mJy, corresponding to a spectral luminosity ($\nu L_{\nu}$) of
$L_{24}=1.38\times{10}^{7}L_{\odot}$. This low luminosity is
consistent with the weak MIR continuum
(Fig. \ref{fig:spectra}b), arising from emission from VSGs heated by the UV radiation field.

We can combine this with a measurement of the H$\alpha$ emission in
this region to obtain a star formation rate (SFR), given that they are
complimentary (H$\alpha$ tracing the young stellar population and
24\micron\ as a measure of dust-absorbed stellar light). We find
$L_{\rm H \alpha}=2.03\times{10}^{6}L_{\odot}$, but caution that 
H$\alpha$ emission in SQ is also the result of shock-excitation
\citep{Xu03} and must be considered an upper limit for measuring star
formation. When we combine $L_{\rm H \alpha}$ with $L_{24}$, using the
relation of \citet{Cal07}, we find a SFR of $<$0.05\,$M_{\sun}\,
\rm{yr^{-1}}$.

A further measure of star formation can be obtained from the PAH strength. Using the relation of \citet{Hou07}, derived from the starburst sample of \citet{Br06},
we can use the 7.7\micron\ flux density as a measure of star formation. The shock subregion has a 7.7\micron\ flux density of 0.73 mJy which corresponds to a SFR of $\sim0.08\,M_{\sun}\,\rm{yr^{-1}}$, in good agreement with our previous calculation, but also an upper limit as some PAH emission is contamination from known star formation regions in the group. A point of caution, however, is that this star formation indicator may be biased given that we detect suppressed 7.7\micron\ emission compared to the 11.3\micron\ PAH. Comparing the $L_{\rm PAH}$(7.7\micron)/$L_{24}$ ratio to the galaxies in SINGS, we find that the value of $\sim0.17$ is typical of star-forming galaxies \citep{Og09}, which suggests that both measures of star formation are low, but self-consistent.

The low {\it upper limits} for the SFR in the shock suggests that star
formation is depressed in the shock apart from in the IGM starburst, SQ-A, which has a SFR of $\sim1.25\,M_{\sun}\,\rm{yr^{-1}}$ at the velocity of the group \citep{Xu03}. This  would be consistent with a picture of
molecular hydrogen being reheated by MHD shocks in the turbulent
medium.% faster than the warm \Ht can cool \citep{Pg09}. 

Since the Jeans' mass increases with gas temperature and turbulence and shearing morions will prevent collapse, the cold molecular gas clouds may be too short-lived or undersized to
facilitate collapse and produce significant star formation \citep{Pg09}. The extent of a cold reservoir of molecular gas in the shock is a key consideration in this scenario, as discussed in Section \ref{CO}.

%Metallicity (\citet{Dal09}:%Centre shock: 9.1

\section{Implications for High-z Galaxy Formation}

\begin{figure*}[!t]
\begin{center}
\includegraphics[width=12cm]{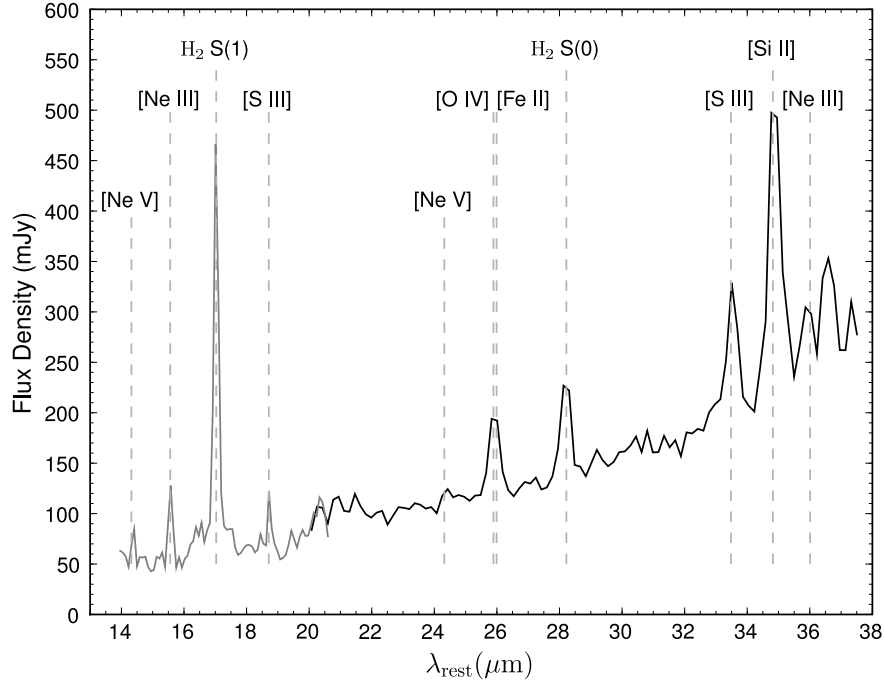}
\caption[]
{\small{Low resolution spectrum of the entire SQ region mapped by the IRS. This extraction is a rectangular box corresponding to an area of 94 kpc $\times$ 56 kpc centered on 22$^{\rm h}$35$^{\rm m}$59.97$^{\rm s}$, +33$\degr$58\arcmin12.7\arcs, and includes the main shock, \Ht bridge, NGC 7319, SQ-A and NGC 7318a and NGC 7318b.}}
\label{fig:full}
\end{center}
\end{figure*}

The present observations suggest that molecular line cooling in
dense clumps dissipates a significant fraction of the kinetic energy available in high-speed shocks.  For reference,
and to provide a glimpse of what this group might look like at high
redshift when filling a single beam, we present in Figure \ref{fig:full}
the spectrum extracted for the entire group (i.e. including the three
neighboring galaxies and the shock).  The total \Ht line
emission from the whole group exceeds ${10}^{42}\, {\rm erg\, s^{-1}}$
and is still the most dominant MIR coolant. Indeed the luminosity of
the rotational \Ht lines is sufficient that it could be detected at
high redshift with future far-infrared or sub-mm instrumentation like
SPICA or SAFIR (see Appleton et al. 2009). How likely is it,
however, that high-speed shocks play a role in the assembly of
galaxies?

There is growing evidence that galaxies at high redshift are turbulent
\citep{For06,Gen08} and increasingly clumpy (e.g. Conselice et
al. 2005, Elmegreen \& Elmegreen 2005). Indeed Bournaud \& Elmegreen
(2009) discuss the importance of the growth instabilities in massive
gas clumps in forming disks at z~$>$~1, and favor at least a large
fraction of the clump systems being formed in smooth flows, perhaps
similar to those discussed by Dekel et al. (2009). To what extent the
build-up of these disks is truly ``smooth'' is not yet clear since the
medium is likely to be a multiphase one. In a more standard
picture, gas flowing into the more massive dark halos will 
experience strong shocks, most likely in an inhomogenious medium
(e. g. Greif et al. 2008) -- thus it begs the question of how
important H$_2$ cooling may be in these different cases.

Models of the collapse of the first structures predict that strong
metal lines soon dominate the cooling over molecular hydrogen when the
first stars pollute the environment. It is therefore usually assumed
that, except at very early stages, molecular hydrogen is a minority
coolant in gas that forms the first major structures (e.g. Bromm et
al. 2001, Santoro and Shull 2006). However, our observations show
that, under the right conditions, even in high metallicity
environments, molecular hydrogen can be extremely powerful -- in this
case dominating by a factor of ten over the usually powerful
$[$Si\,{\sc ii}$]34.82 \mu$m line in strong shocks. If there are
situations at high redshift where strong shocks propagate into a
clumpy, multiphase medium, then our observations imply that molecular
hydrogen cooling cannot be assumed to be negligible. On the other
hand, this will not be a trivial problem. Our best model of Stephan's
Quintet \citep{Pg09} involves the formation of \Ht in a complex
multiphase turbulent medium in which shocks destroy dust in some
places, but allow survival in others - thus encouraging \Ht formation.

In the early universe, this enhanced cooling, which has so far been neglected, will depend on the distribution and nature of the first dust grains, in concert with the formation, temperature and abundance of gas, and the feedback effects from the first stars and AGN.

%The compact group Stephan's Quintet provides an ideal laboratory for
%studying the tidal interactions that may be analogous to the formation
%of massive galaxies at $z \sim 1-2$. The group galaxies have undergone
%vigorous \HI stripping, this material now being part of the IGM which
%is currently undergoing shock heating as the result of a galaxy
%collision. If this system is representative of the mergers common at
%high redshift that produced massive galaxies, the shock-heating of
%molecular hydrogen can represent an observable in terms of probing
%early galaxy formation.

%In Figure \ref{fig:full} we show an extraction of the entire region
%around the shock, including the three neighbouring galaxies that are
%part of the group. We see signatures of star formation as well as
%high-excitation due to the AGN in NGC 7319. However, it is striking
%that the \Ht emission still dominates the spectrum. The total \Ht line
%emission from the whole group exceeds ${10}^{42}\, {\rm erg\, s^{-1}}$.

%Current models of the formation and excitation of \Ht in these
%galaxy-wide collisions have important implications for the energetics
%of galaxy formation, as well as other shocked systems in the universe.

\section{Conclusion}

In this paper we have presented the results of the mid-infrared spectral mapping of the Stephan's Quintet system using the {\it Spitzer Space Telescope}. We highlight here our five main conclusions:

\begin{itemize}

\item{The powerful \Ht emission detected by \citet{App06} surprisingly represents only a small fraction of the group-wide warm \Ht (with a lower limit luminosity of $1.1\times {10}^{42}\, {\rm erg\, s^{-1}}$ spread over $\sim975$\, kpc$^2$) that dominates the mid-infrared emission of the system. There is evidence for another shock-excited feature, the so-called \Ht bridge between the main shock and NGC 7319, which is likely a remnant of past tidal interactions within the group. The spatial variation in the distribution of the \Ht 0-0 line ratios implies differences in temperature and excitation in the shocked system - this will be explored fully in Paper II.}

\item{The global L(H$_{2}$)/L$_{X(0.001-10 \rm kev)}$ ratio in the main shock is $>= 3$, and $\sim$2.5 in the new ``bridge'' feature. The results confirm that MIR H$_2$ lines are a stronger coolant than X-ray emission over the shock structures, indicating a new cooling pathway seen on a large scale in SQ. This modifies the traditional view that X-rays dominate cooling at all times in the later stages of compact group evolution. Since \Ht forms on the surfaces of dust grains, we expect dust emission associated with these regions, but a low intensity radiation field produces only weak emission at 24\micron.}

\item{Following earlier interpretations of nebular line ratios in the optical, we interpret infrared ionic lines within the framework of fast ($V_{s} > 100$ \,km\,$\rm{s^{-1}}$) ionizing shocks. Comparison between the $[$Ne\,{\sc ii}$]$, $[$Ne\,{\sc iii}$]$, $[$Si\,{\sc ii}$]$ and $[$Fe\,{\sc ii}$]$ line intensities implies that both silicon and iron are depleted onto dust. This result implies that dust is not destroyed in the shock.%emission associated with the shock is consistent with regions of fast shocks ($100 <V_{s} < 300$\,km\,$\rm{s^{-1}}$) experiencing depletion onto dust.
}

%of collisional heating in dense clouds associated with the molecular gas. In such clouds it is relatively easy to generate strong $[$Si\,{\sc ii}$]$ emission without the need for significant depletion of Silicon from grains.}

%\item{Investigation of the star formation properties associated with the shock region have shown nominal star formation, surprising given the large reservoir of molecular hydrogen present. Suppression of star formation could be due to turbulent conditions continuously heating the \Ht and as a result regions of cool \Ht are either short-lived or not massive enough to collapse, thus quenching star formation.}

\item{Star formation in SQ is dominated by SQ-A and 7318b-south, located at the extreme ends of the shock ridge seen at radio wavelengths, suggesting they are both shock triggered starbursts. However, regions dominated by warm \Ht emission exhibit very low star formation rates, consistent with a turbulent model where \Ht is significantly reheated and cool clouds are too short-lived or undersized to collapse.}

\item{In SQ we observe the projected coexistence of $[$Si\,{\sc ii}$]$ and H$_{2}$ being produced by $\sim$200\,km\,$\rm{s^{-1}}$ and $\sim$20\,km\,$\rm{s^{-1}}$ velocity shocks, respectively. Our observational results are consistent with a model of a multiphase postshock medium produced by a galaxy-wide collision \citep{Pg09}.}

\end{itemize}

The cooling pathway of warm \Ht emission we observe group-wide in SQ is clearly a significant, albeit surprising, mechanism in shock systems. %It is plausible that the diverse environments producing strong warm \Ht emission have this in common. 
To determine the overall dominant cooling mechanism in SQ, we require an inventory of lines and continuum processes at all wavelengths. Early shock models \citep{Dr83} predict that, apart from the rotational emission from H$_{2}$, contributions from lines such as $[$O\,{\sc i}$]$63.2\micron, $[$C\,{\sc ii}$]$157.7\micron\ and the THz spectrum of H$_{2}$0 \citep{Pear91} could be significant. We hope to explore this chemistry more fully, and the detailed distribution of cool dust, using the capabilities of the {\it Herschel} Space Observatory. In addition, we cannot rule out strong UV line-cooling.

Stephan's Quintet provides the ideal laboratory for probing a mechanism potentially crucial in systems ranging from ULIRGs to radio galaxies to supernova remnants.

\acknowledgements

MEC is supported by NASA through an award issued by JPL/Caltech under Program 40142. We thank Tom Jarrett for use of his IRS pixel cleaning software and IRAC/MIPS photometry software.

\newpage

\section*{Appendix A: The Seyfert II Galaxy, NGC 7319}\label{AGN}

In this Section we discuss the results pertaining to NGC 7319, a Seyfert 2 galaxy \citep{Dur94} lying to the East in the SQ group (see Fig. 1).
The specific intensity contours for the H$_{2}$ S(0) and H$_{2}$ S(1) lines (Fig. 2a and b) show emission
associated with the galaxy, seemingly connected to the rest of the group by the \Ht ``bridge'' discussed in Section \ref{molhy}.
Figure \ref{fig:xray_rad} shows that the nucleus of NGC 7319 produces strong X-ray emission \citep{Trin03} and is prominent at radio wavelengths \citep{Xu03}. However, we observe an offset between the peak of the \Ht emission near NGC 7319 and the Seyfert nucleus, which suggests that the ``bridge'' is a separate structure and not being excited by the AGN.

The AGN in NGC 7319 does not have a well-collimated jet, but two extended lobes with compact hotspots, asymetrically distributed along the minor axis of the galaxy \citep{Xan04}. This structure runs NE/SW and its orientation compared to the \Ht filament (which runs EW) is not consistent with causing the excitation of the \Ht bridge. This is also evident from the relatively weak power in the AGN, as inferred by the emission line diagnostics discussed below, and in the X-ray where it is only a factor of $\sim$2 greater than the emission associated with the main shock and ``bridge''. 
%\citep{Og07} investigated shock heating of \Ht by radio jets in 3C 326, but found that not enough kinetic energy could be dissipated by the jet to heat the \Ht.

In Figure \ref{fig:fs} we present the specific intensity contours of the $[$Fe\,{\sc ii}$]$25.99\micron\ (blended with $[$O\,{\sc iv}$]$25.89$\mu$m), $[$S\,{\sc iii}$]$33.48\micron\ and
$[$Si\,{\sc ii}$]$34.82\micron\ emission lines. 
Given the low spectral resolution of the SL and LL modules of {\it Spitzer}, 
we cannot distinguish between emission from $[$Fe\, {\sc ii}$]$ and 
$[$O\,{\sc iv}$]$. The AGN in NGC 7319 is likely to
produce both, with $[$Fe\,{\sc ii}$]$ emission likely originating in
X-ray dissociation regions (XDRs) surrounding the AGN
\citep{Mal96}. %We discuss specifically the mid-IR spectrum of NGC 7319 below.
Prominent emission from $[$S\,{\sc iii}$]$33.48\micron\ and
$[$Si\,{\sc ii}$]$34.82\micron\ are due to the high excitation conditions associated with the AGN. Dense PDRs and X-ray dominated regions, powered by AGN, show strong $[$Si\,{\sc ii}$]$ emission at 34.82\micron, while $[$S\,{\sc iii}$]$ 33.48 \micron\ emission acts as a tracer of \HII\ regions.

\begin{figure*}[!th]
\begin{center}
\subfigure[]{\includegraphics[width=5.5cm]{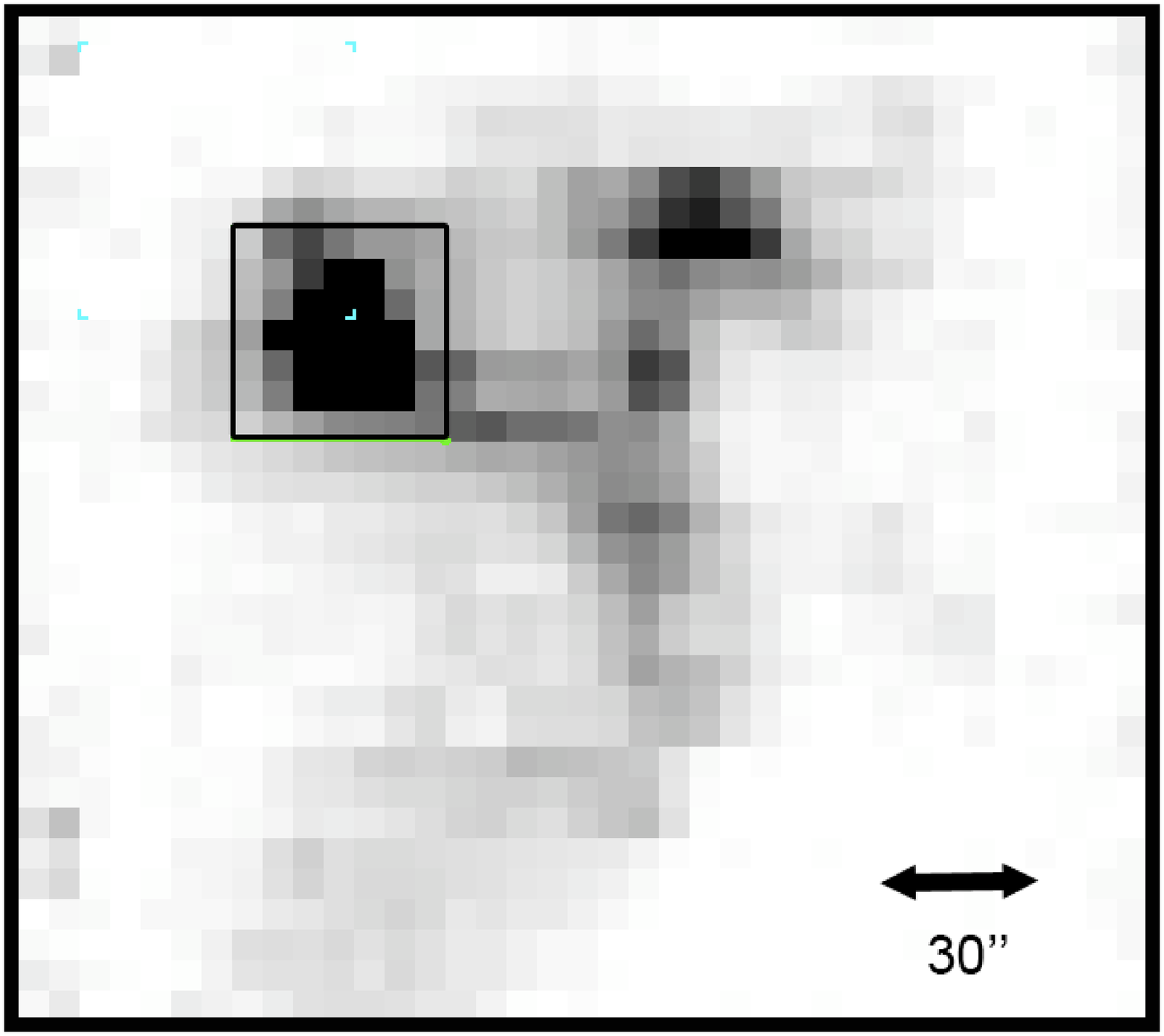}}
\hfill
\subfigure[]{\includegraphics[width=10.5cm]{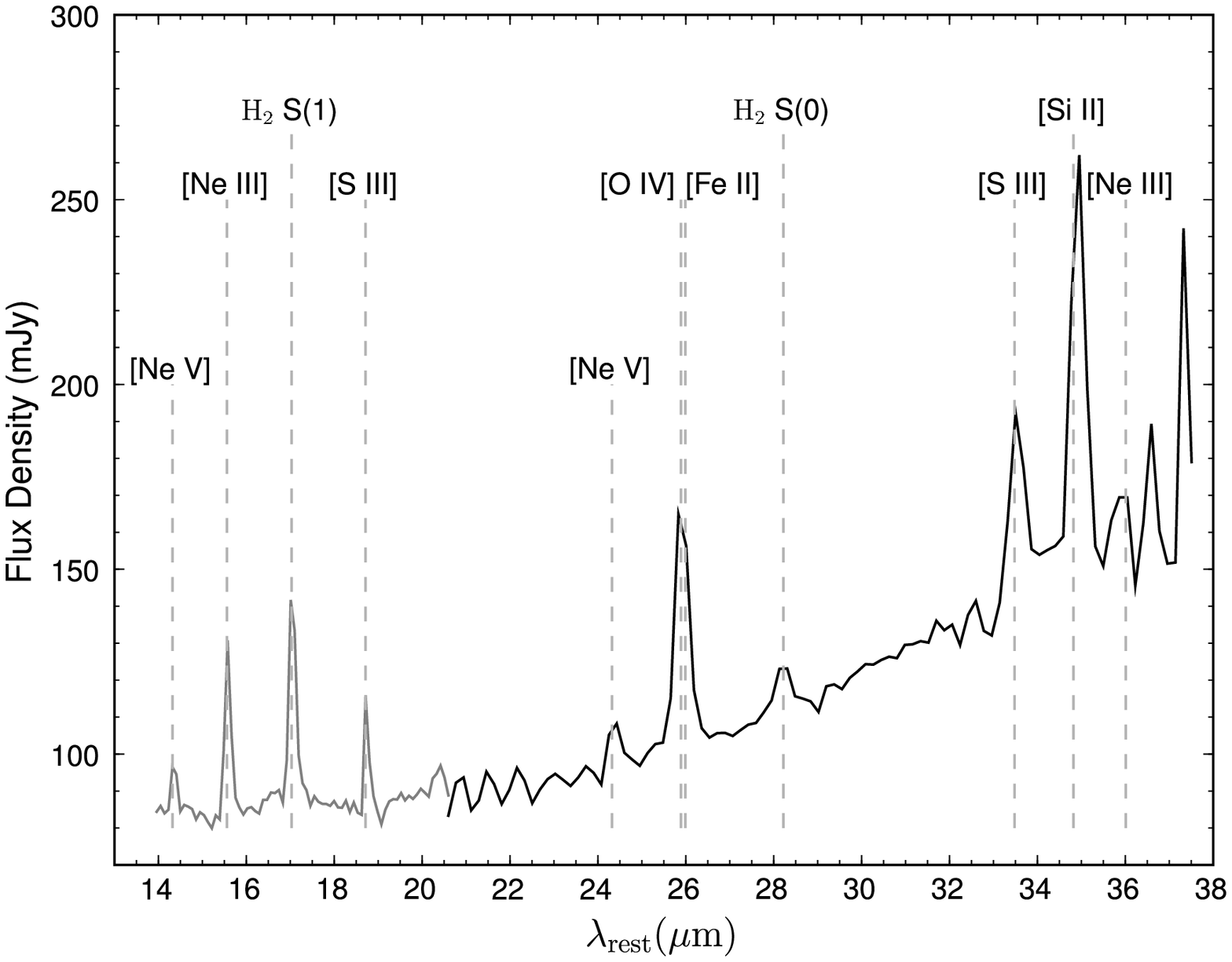}}
\caption[]
{\small{Rectangular extraction box for the Seyfert 2 galaxy NGC~7319 overlaid on the \Ht 0-0 S(1) line emission (a) and the corresponding low-resolution spectrum (b). The extraction box is centered on 22$^{\rm h}$36$^{\rm m}$03.62$^{\rm s}$, +33$\degr$58\arcmin35.6\arcs. }}
\label{fig:AGN}
\end{center}
\end{figure*}

We now focus on the emission line properties of NGC 7319. Figure \ref{fig:AGN} shows the spectrum extracted from the galaxy and the measured properties of the line ratios are listed in Table \ref{tableh2fluxes} and \ref{tablemetalfluxes}. For the first time in the SQ group, we observe a spectrum
which is no longer dominated by \Ht emission, but instead the brightest lines
are the high-excitation $[$O\,{\sc iv}$]$25.91$\mu$m and $[$Fe\,{\sc ii}$]$25.98\micron\ blended
lines, as well as the $[$Si\,{\sc ii}$]$34.81$\mu$m line. $[$Ne\,{\sc v}$]$ is prominent at both
14.32\micron\ and 24.30$\mu$m -- a line typically seen in AGN. Also, unlike the
majority of the extended shocked regions, there is a rising thermal
continuum more typical of a starforming galaxy than a classical Seyfert galaxy,
although as various studies have shown (Buchanan et al. 2006 ; Deo et
al. 2007), Seyfert II galaxies exhibit a variety of MIR
spectral characteristics at long wavelengths. NGC~7319's rising
continuum is similar to that seen in the Seyfert II galaxy NGC~3079
(Deo et al. 2007), and likely represents a dominant starburst
component in the Far-IR. Spitzer imaging reveals not only a bright
nucleus, but also extended emission regions in the galaxy.
% (more needed? Tie in with CO imaging and H2?)

Using the $[$Si\,{\sc ii}$]$34.81$\mu$m/$[$S\,{\sc iii}$]$33.48\micron\ ratio as a probe of excitation sources, we find a ratio of $\sim$1.85 which is low compared to the average value found for AGN galaxies ($\sim 2.9$) in the {\it Spitzer} Infrared Nearby Galaxy Sample (SINGS), but high compared to star-forming regions ($\sim 1.2$) in the same sample \citep{Dal06}. This suggests a relatively weak AGN. The $[$Ne\,{\sc iii}$]$\,15.56\micron/$[$Ne\,{\sc ii}$]$\,12.81\micron\ ratio is a measure of radiation field strength and the value of 0.97 indicates a typical radiation field strength compared to other AGN (the sample of Weedman et al. 2005 shows a range of $\sim$0.17 to 1.9). The average electron density, estimated from the
$[$S\,{\sc iii}$]$18.71\micron/$[$S\,{\sc iii}$]$33.48\micron\ ratio of $\sim$0.56, is $100-200$ cm$^{-3}$ \citep{Mar02}, in the low-density limit
for this diagnostic \citep{Smi09}.

%\begin{figure*}[!tbh]
%\begin{center}
%\hfill
%\includegraphics[width=12cm]{fig15.eps}
%\caption[]
%{\small{Warm \Ht Excitation Diagram for NGC 7319. The ordinate (N/g) represents the column density divided by the statistical weight of the rotational state. }}
%\label{fig:excite7319}
%\end{center}
%\end{figure*}

The properties of the \Ht emission of NGC~7319 are limited to the two
long wavelength lines observed by IRS-LL (the galaxy lies outside the
region mapped by SL). As a result, the excitation diagram contains
only two points allowing only an approximate idea of the \Ht mass,
since without the SL wavelength coverage to provide information on
possible warmer components, we are likely to overestimate the
temperature of the \Ht by fitting a straight line to the points.
%(shown in Fig. \ref{fig:excite7319}). 
Any warmer component would contribute
to the flux of the 0-0 S(1) line thus leading to a reduction of the
temperature (and increase in \Ht mass) for any cooler
component. However, to provide a guide, we estimate the temperature of
\Ht $<=$171$\pm 8$ K (assuming the gas is in thermal equilibrium and thus
an ortho-para ratio of 2.65) and a total \Ht mass of
3.0$\pm 7 \times$10$^{8}$\,M$_{\odot}$.  We consider this a lower limit to the
total warm mass if (as is likely) more than one component is present.  Gao \& Xu (2000)
estimated the {\it cold} molecular hydrogen mass of NGC 7319 based on $^{12}$CO (1-0) observations
as $>$ 3.6~$\times$10$^9$\,M$_{\odot}$, a factor of roughly ten greater than the warm \Ht  mass.
Such a ratio is not atypical of large spiral galaxies \citet{Rig02}.
%, although it is perhaps surprising that
%the ratio is relatively normal, given how abnormal the gas distribution is in SQ!

\section*{Appendix B: High Resolution Spectrum of Shock Centre Reanalysed}\label{SQ_HR}

\begin{figure*}[!th]
\begin{center}
\includegraphics[width=16cm]{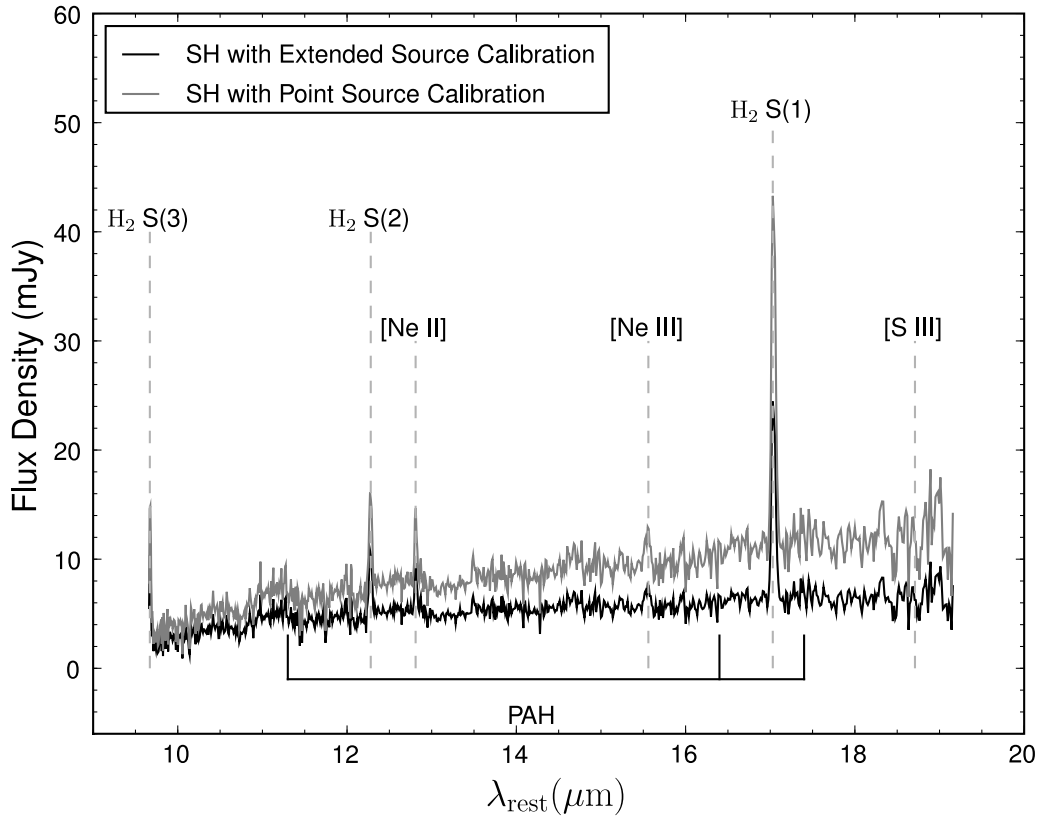}
\caption[]
{\small{{\it Spitzer} Short-High (SH) Spectrum, centred on 22$^{\rm h}$35$^{\rm m}$59.57$^{\rm s}$, +33$\degr$58\arcmin1.8\arcs, reanalysed using the latest point source and extended source calibrations. }}
\label{fig:HR}
\end{center}
\end{figure*}

The high resolution spectrum at the center of the shock in SQ, obtained by \citet{App06}, has been reanalysed using the latest calibrations available for the IRS instrument (SSC pipeline version S17) and is included as Figure \ref{fig:HR}. This has shown that the \Ht gas lies at a velocity of 6360($\pm$100) km s$^{-1}$, between the velocity of the group (6600 km\,s$^{-1}$) and the velocity of the intruder (5700 km\,s$^{-1}$). This is consistent with a model of gas being accelerated by the shock, as well as the turbulence demonstrated by the broad linewidth of the \Ht in the shock (860 km\,s$^{-1}$).

The new SH spectrum shows that with the improved calibrations, the 11.3\micron\ PAH feature is detected, although faint. This is consistent with what is found in the larger extractions that show enhanced 11.3\micron\ PAH emission compared to ionised PAH features emitting at 6.2\micron\ and 7.7\micron.

\section*{Appendix C: X-ray Fluxes in Extracted Regions}\label{xray}

%In Appendix C, we present a complete reanalysis of the XMM-Newton
%observations of SQ using the latest calibrations, the original data of
%which was presented by \citet{Trin05}. The reason for the reanalysis from
%archival data was to obtain accurate fluxes and luminosities for the
%X-ray emission over the various apertures matched to our spectral
%extractions. The results of the analysis are presented in Table A1 and
%matches exactly the extraction results of Table~\ref{tableh2fluxes}
%and Table~\ref{tablemetalfluxes}, and are in
%good agreement with \citet{Trin05} in terms of overall
%properties of these data.

In this section we present the X-ray fluxes in the extraction regions shown in Figure \ref{fig:ext}. The reanalysis of archival data is necessary to obtain accurate fluxes and luminosities for the
X-ray emission over the various apertures matched to our spectral
extractions. We use the XMM-Newton EPIC-pn data \citep[see][for observational details]{Trin05} to obtain the most sensitive measurements. 
A calibrated event file was generated and filtered using standard
 quality flags, and subsequently cleaned of background flares.  A
 0.3--2~keV image was then extracted and corrected for instrumental
 response. All point-like sources were masked out to $25\arcsec$ in
 the analysis of diffuse emission, and the local background level was
 estimated within a $5\times 9$~arcmin$^2$ rectangular region away
 from the group core. For each region in Fig.~7, resulting
 background-subtracted photon count rates were converted to
 0.3--2~keV and ``bolometric" (0.001--10~keV) X-ray fluxes assuming
 an absorbed thermal plasma model of metallicity 0.4 Solar, with
 temperature as estimated from the map of O'Sullivan et al. (2009),
 and an absorbing Galactic H{\sc i} column density of $N_{\rm
 H}=6.2\times 10^{20}$~cm$^{-2}$.  For NGC\,7319, which harbors an
 AGN, an absorbed power-law spectrum of photon index $\Gamma=1.7$ was
 assumed instead.
Results are listed in Table \ref{X-ray}.

\begin{table}[!ht]
{\scriptsize
\begin{minipage}{160mm}
\begin{center}
\caption{X-ray Count Rates and Fluxes for Extracted Regions}
%\centering
\begin{tabular}{lcccc}
\tableline\tableline
\\[0.25pt]
Region & Count Rate (0.3-2 keV) & Flux (0.3-2 keV) & Bolometric Flux (0.001-10 keV) & Adopted Spectral Model \\  
            & (photons\,s$^-1$) & erg\,s$^{-1}$\,cm$^{-2}$ &   erg\,s$^{-1}$\,cm$^{-2}$ &   \\
\tableline
\\[0.25pt]

Main Shock   & 0.102  & 1.8 $\times 10^{-13}$  & 2.8 $\times 10^{-13}$ &  T = 0.7 keV\\
Shock Sub-region  & 0.017  & 3.3 $\times 10^{-14}$  & 5.1 $\times 10^{-14}$ &  T = 0.8 keV\\
Bridge       & 0.017  & 3.0 $\times 10^{-14}$  & 4.5 $\times 10^{-14}$ &  T = 0.6 keV\\
NGC~7319     & 0.051  & 1.1 $\times 10^{-13}$  & 3.7 $\times 10^{-13}$ &  $\Gamma$ = 1.7\\
SQ-A         & 0.013  & 2.4 $\times 10^{-14}$  & 3.7 $\times 10^{-14}$ &  T = 0.6 keV\\
\tableline

\end{tabular}
\tablecomments{Statistical uncertainties on all flux measurements are $<10$\%.}.

\label{X-ray}
\end{center}
\end{minipage}
}
\end{table}

We note that for the main shock (with luminosity 1.9~$\times$10$^{41}$\,erg\,s$^{-1}$ in the 0-3-2 keV band) we are within a factor of 2 of the 0.5-2 keV luminosity $\sim$3.1~$\times$10$^{41}$\,erg\,s$^{-1}$ obtained for a similar, but larger extraction of the shock by \citet{Trin05}. \citet{Osul09} obtain a 0.5-2 keV surface brightness of 0.07 L$_{\odot}$\,pc$^{-2}$ in the main shock compared to our value of 0.1 L$_{\odot}$\,pc$^{-2}$ for the 0.3-2 keV surface brightness.

\clearpage

\begin{table}

%{\bf [Table contains extra extraction region (shock center) fluxes for now]}

{\scriptsize
\caption{Observed H$_2$ Line Fluxes in units of 10$^{-17}$~W m$^{-2}$}
%\centering
\begin{tabular}{lccccccc}
\tableline\tableline
\\[0.25pt]
Target Region & Aperture & H$_{2}$ 0-0 S(0) & H$_{2}$ 0-0 S(1) & H$_{2}$ 0-0 S(2) & H$_{2}$ 0-0 S(3) & H$_{2}$ 0-0 S(4) & 
H$_{2}$ 0-0 S(5) \\  
              & (arcsec$^2$) & $\lambda$28.22$\mu$m              & $\lambda$17.03$\mu$m                   & $\lambda$12.28$\mu$m     
             & 
$\lambda$9.66$\mu$m                  & $\lambda$8.03$\mu$m                  & $\lambda$6.91$\mu$m  \\
\tableline
\\[0.25pt]
%                       S(0)          S(1)                S(2)           S(3)           S(4)          S(5)
Main Shock   & 2307 & 3.09$\pm$0.19 & 23.05$\pm$0.26 & 9.10$\pm$0.38 & 22.76$\pm$0.84 & 2.5$\pm$1.0  & 14.1$\pm$0.7  \\
Shock Sub-region  & 242  & 0.36$\pm$0.03& 3.13$\pm$0.02  & 1.47$\pm$0.08 & 3.60$\pm$0.08  & 1.04$\pm$0.15 & 2.37$\pm$0.30  \\ 
%Shock Center & 274  & 0.40$\pm$0.05& 3.58$\pm$0.03  & 1.52$\pm$0.05 & 4.10$\pm$0.09  & 0.96$\pm$0.12 & 2.68$\pm$0.22  \\ 
Bridge       & 413  & 0.72$\pm$0.04 &  4.94$\pm$0.09 & \nodata       & \nodata        & \nodata   & \nodata       \\
NGC~7319     & 1302 & 2.06$\pm$0.26 & 11.00$\pm$0.48 &  \nodata      & \nodata        & \nodata   & \nodata       \\
SQ-A         & 671  & 1.83$\pm$0.05 & 8.8$\pm$0.13   & \nodata       & \nodata        & \nodata   & 3.7$\pm$0.93  \\  
\tableline
\end{tabular}
\tablecomments{Upper limits are the 3$\sigma$ values calculated from the RMS (root mean square) and the expected line profile width.}.
\label{tableh2fluxes}
}
\end{table}

\begin{table}
{\scriptsize
%{\bf [Need to add PAH measurements and luminosities]}

\caption{Observed Fine Structure Line Fluxes in units of 10$^{-17}$~W m$^{-2}$}
%\centering
\begin{tabular}{lcccccccc}
\tableline\tableline
\\[0.25pt]
Target Region  & $[$Ne\,{\sc ii}$]$ & $[$Ne\,{\sc v}$]$ & $[$Ne\,{\sc iii}$]$ & $[$S\,{\sc iii}$]$ & $[$Ne\,{\sc v}$]$& $[$Fe\,{\sc ii}$]$+$[$O\,{\sc iv}$]$ & $[$S\,{\sc iii}$]$     & $[$Si\,{\sc ii}$]$     \\  
              &$\lambda$12.81$\mu$m &$\lambda$14.32$\mu$m &$\lambda$15.56$\mu$m &$\lambda$18.71$\mu$m &$\lambda$24.32$\mu$m & $\lambda$25.99+25.89$\mu$m & $\lambda$33.48$\mu$m  &$\lambda$34.82$\mu$m  \\
\tableline
\\[0.25pt]
%                       [NeII]          [NeV]           NeIII          SIII            NeV          FeII/OIV       SIII            SiII
Main Shock    & 5.31$\pm$0.30 & $<$0.9       & 1.94$\pm$0.39 &    0.76$\pm$0.13        & $<$0.14    & 1.07$\pm$0.10  & 1.87$\pm$0.15
 & 8.59$\pm$0.27         \\
Shock Sub-region  & 1.04$\pm$0.09 & $<$0.18     & 0.15$\pm$0.06 & $<$0.12        & $<$0.14   & 0.15$\pm$0.03        & 0.13$\pm$0.02             &
 1.23$\pm$0.04         \\ 
%Center Center & 274  & 1.1$\pm$0.06 & $<$0.16     & 0.31$\pm$0.05 & $<$0.12        & $<$0.14   & 0.18$\pm$0.03        & 0.19$\pm$0.02             &
 %1.30$\pm$0.04         \\ 
Bridge       & \nodata       & \nodata        & 0.39$\pm$0.04 & $<$0.1        & $<$0.15     & 0.29$\pm$0.04 &  $<$0.22    &  1.1$\pm$0.1     
    \\
NGC~7319      & \nodata  & 2.89$\pm$0.64  &  9.47$\pm$0.29& 2.81$\pm$0.40 & 1.56$\pm$0.39 & 10.9$\pm$0.27 & 5.0$\pm$0.42 & 9.24
$\pm$0.49      
     \\
SQ-A          & \nodata & $<$0.8   & 1.29$\pm$0.16  & 0.87$\pm$0.06 & $<$0.18   & $<$0.17  & 1.54$\pm$0.10& 2.3$\pm$0.16
   \\  
\tableline
\end{tabular}
\tablecomments{Upper limits are the 3$\sigma$ values calculated from the RMS and the expected line profile width.}.
\label{tablemetalfluxes}
}
\end{table}

\end{document}